\documentclass[fleqn,usenatbib]{mnras}
\usepackage{graphicx}
\usepackage{amsmath}
\usepackage{amssymb}
\usepackage{natbib}
\usepackage{cases}
\usepackage{paralist}
\usepackage{tablefootnote}
\usepackage{bm}
\usepackage{upgreek}
\usepackage{bbold}
\usepackage{url}
\usepackage{hyperref}
\newcommand{\taurex}{TauREx }
\newcommand{\nemesis}{NEMESIS }
\newcommand{\chimera}{CHIMERA }

\DeclareMathAlphabet{\mathpzc}{OT1}{pzc}{m}{it}

\title[Retrieval Comparison]{A comparison of exoplanet spectroscopic retrieval tools}

\author[Barstow, Changeat, Garland, Line, Rocchetto, Waldmann]{Joanna K. Barstow,$^{1}$\thanks{j.barstow@ucl.ac.uk}, Quentin Changeat$^{1}$, Ryan Garland$^{2}$, Michael R. Line$^{3}$, \newauthor{Marco Rocchetto$^{1}$, Ingo P. Waldmann$^{1}$}
\\
$^{1}$Department of Physics and Astronomy, University College London, Gower Street, London, WC1E 6BT\\
$^{2}$Atmospheric, Oceanic and Planetary Physics, Clarendon Laboratory, University of Oxford, Keble Road, Oxford, OX1 3PU\\
$^{3}$School of Earth and Space Exploration, Arizona State University, PO Box 871404. Tempe, AZ 85281.
}

\date{Accepted XXX. Received YYY; in original form ZZZ}

\pubyear{2019}

\begin{document}
\label{firstpage}
\pagerange{\pageref{firstpage}--\pageref{lastpage}}
\maketitle

\begin{abstract}
Over the last several years, spectroscopic observations of transiting exoplanets have begun to uncover information about their atmospheres, including atmospheric composition and indications of the presence of clouds and hazes. Spectral retrieval is the leading technique for interpretation of transmission spectra and is employed by several teams using a variety of forward models and parameter estimation algorithms. However, different model suites have mostly been used in isolation and so it is unknown whether the results from each are comparable. As we approach the launch of the James Webb Space Telescope we anticipate advances in wavelength coverage, precision, and resolution of transit spectroscopic data, so it is important that the tools that will be used to interpret these information rich spectra are validated. To this end, we present an inter-model comparison of three retrieval suites: TauREx , NEMESIS and CHIMERA. We demonstrate that the forward model spectra are in good agreement (residual deviations on the order of 20$—-$40 ppm),  and discuss the results of cross retrievals between the three tools. Generally, the constraints from the cross-retrievals are consistent with each other and with input values to within 1$\sigma$. However, for high precision scenarios with error envelopes of order 30 ppm, subtle differences in the simulated spectra result in discrepancies between the different retrieval suites, and inaccuracies in retrieved values of several $\sigma$. This can be considered analogous to substantial systematic/astrophysical noise in a real observation, or errors/omissions in a forward model such as molecular linelist incompleteness or missing absorbers.
\end{abstract}


\section{Introduction}
Recently, the field of exoplanet atmospheres has undergone dramatic expansion, with spectroscopic observations now available for several tens of transiting planets. When a planet passes in front of its parent star during its orbit, some of the starlight is filtered through the planet's atmosphere, and spectroscopic measurements of the transit depth can reveal details of atmospheric composition and structure. The majority of objects studied in this way are highly-irradiated hot Jupiters - planets similar in size to solar system gas giants, but in extremely close orbits around their parent stars (e.g. \citealt{sing16}). Other intriguing planets include super Earths such as GJ 1214b \citep{irwinj09,kreidberg14} and 55 Cancri e \citep{fischer08,tsiaras16} - planets with no solar system analog, lying between the Earth and Neptune in size and occupying the transitional region between terrestrial planets and gas giants. 

So far, the Space Telescope Imaging Spectrograph and Wide Field Camera 3 on the \textit{Hubble Space Telescope} (\textit{HST}) and the InfraRed Array Camera on \textit{Spitzer} have been the workhorse instruments for transiting exoplanet science. Recently, data quality has increased to the point of meaningful comparison between spectra of different planets (e.g. \citealt{sing16}, \citealt{stevenson14b}). The launch of the \textit{James Webb Space Telescope} (\textit{JWST}) in 2021 will provide spectra of large signal-to-noise, spectral resolution, and wavelength coverage and thus increase our capacity for comparative exoplanet atmospheric science \citep{barstow15,greene16,morley17}. For the first time, progress in our understanding of exoplanet atmospheres is likely to be be limited by model completeness and robustness rather than data quality. 

Atmospheric retrieval has been the tool of choice for the interpretation of transiting exoplanet spectra. Retrieval algorithms generate (usually a 1-D atmosphere) a forward model spectra of exoplanet atmospheres, then iteratively solve the inverse problem to find the best fitting model solution to the observed data. This technique has been used extensively on both real (e.g. \citealt{madhu09,Madhu11, line13a,kreidberg14, fraine14,madhu14,barstow17,tsiaras18,fisher18}) and simulated data sets (\citealt{benneke12, greene16, waldmann14, Molliere19, Blecic17, Zhang19}) and is generally acknowledged \citep{NAS18, madhu18} to be an efficient and reliable method for constraining exoplanet atmospheres from transmission and eclipse spectra. 

Whilst all retrieval codes follow the same basic structure: a forward model that produces a spectrum, the data, and a Bayesian parameter estimator (e.g., see Figure 4 in \cite{MacMadhu17} ), there is substantial variation in both the forward model details and the methods used to determine the parameter constraints/posterior distributions. This leads to the possibility that two different retrieval tools may provide vastly different solutions to the same dataset, which is clearly not an ideal situation. For example, analyses of the hot Jupiter WASP-63b undertaken by four separate teams are presented by \citet{kilpatrick17}, and some substantial discrepancies are apparent between the solutions for H$_2$O abundance in particular. By contrast, independent retrievals of the transmission spectrum of WASP-52b presented by \cite{bruno19} produce more consistent results. In order to understand the underlying reasons for discrepancies between retrieval results from the same data set, and to avoid occurrences of similar problems in the \textit{JWST} era, a direct comparison of retrieval algorithms is warranted.  We note, an equally important investigation of self-consistent 1D forward models (\cite{Baudino17}) found, unsurprisingly, that differences in the opacity source choices/assumptions resulted in the largest model discrepancies.

To test the robustness of model differences within atmospheric retrievals, we here present a comparison of three different retrieval codes which have all been previously used to analyse transmission spectra of exoplanets.~\nemesis is an optimal estimation retrieval algorithm originally developed for solar system planets \citep{irwin08}, which has recently been upgraded to also incorporate a nested sampling algorithm \citep{krissansen-totton18}.~\taurex \citep{waldmann15,Waldmann16} and~\chimera \citep{line13a} were both developed for application to exoplanet spectra and also use a nested sampling algorithm.  We would also like to emphasise that numerous ``model upgrades" can happen over the course of multiple publications/years. It is therefore expected that such tools are always likely in a continuous evolution, attempting to integrate newer parameterizations, opacities, or sampling methods.  Ergo, we provide details of the current versions of these three specific tools used in this work, in Section~\ref{models} below.

\section{Model Descriptions}
All three models are set up to include the same parameters for this study. These are the constant-with-altitude abundances for H$_2$O, NH$_3$, CO$_2$, CH$_4$, and CO, the planet radius at 10 bar, an isothermal``scale height" temperature, and a hard grey ``cloud top" pressure. The remaining gas is assumed to be H$_2$/He (at approximately solar proportions, 0.85:0.15), with the exception of a high mean molecular weight example, for which N$_2$ is also included. We note also that all three routines utilise the correlated-K treatment of opacities \citep{lacis91}, assuming a Voigt profile shape with a line wing cutoff at 25 cm$^{-1}$. The calculation of transmission through the terminator of the planet follows closely the method described by \cite{tinetti12} for all routines; all codes use the PyMultiNest sampler \citep{buchner14,feroz09} to explore the parameter space.    

\label{models}

\subsection{NEMESIS}
The~\nemesis (Non-linear optimal Estimator for MultivariatE spectral analySIS; \citealt{irwin08}) retrieval algorithm was originally developed to analyse data from the \textit{CASSINI} spacecraft orbiting Saturn (e.g. \citealt{fletcher09}). It was later extended to work for any Solar System planet (e.g. \citealt{barstow12}), and later modified for exoplanets \citep{lee12}.~\nemesis incorporates a fast correlated-k radiative transfer model and options for nested sampling or optimal estimation retrieval algorithms. The correlated-k approximation is a way of pre-tabulating gas absorption coefficients within a wavelength interval, relying on the assumption that the strongest lines at one level in the atmosphere are correlated with the strongest lines at other levels. For this work, we use 20 Gaussian quadrature points to sample the distribution within each spectral interval. Further details may be found in \citet{irwin08}, \citet{goodyyung} and \citet{lacis91}. NEMESIS also includes collision-induced absorption and Rayleigh scattering due to H$_2$ and He. Collision-induced absorption is obtained from the HITRAN12 database \citep{richard12}. H$_2$-broadening is included according to the same prescription used by \citet{amundsen14}. 

Whilst NEMESIS was originally developed to work with a fast and efficient optimal estimation algorithm, a disadvantage was that optimal estimation does not allow full exploration of non-Gaussian posterior distributions. NEMESIS has therefore been recently upgraded to also include the PyMultiNest algorithm\citep{krissansen-totton18}. 

\subsection{TauRex}
TauREx\footnote{\url{https://github.com/ucl-exoplanets/TauREx3_public}} (Tau Retrieval for Exoplanets) is a fully Bayesian radiative  transfer  and  retrieval framework. TauREx can be used with the line-by-line cross sections from the Exomol project \citep{tennyson16} and HITEMP \citep{rothman14} and HITRAN \citep{gordon16}. Temperature and pressure dependent line-broadening was included, taking into account J-dependence where available \citep{Pine1992}.  The line-by-line cross sections can be converted to the correlated-k tables and in this paper, we used the correlated k-tables for H$_2$O, NH$_3$, CH$_4$, CO and CO$_2$ at a fixed resolving power of 100, assuming 20 Gauss-quadrature points per wavenumber bin. While we only investigate the primary transit situations here, TauREx can be used in both transmission spectroscopy model \citep{waldmann15} and emission from secondary transits \citep{Waldmann16}. We also included absorptions from Rayleigh scattering and CIA for the couples H$_2$-H$_2$ and H$_2$-He \citep{Borysow2001,borysow02, Rothman2013}.  The public version of TauREx is able to retrieve chemical composition  of  exoplanets  by  assuming  constant abundances with altitude or equilibrium chemistry \citep{venot01}. To perform the retrieval, TauREx uses the nested sampling retrieval algorithm Multinest \citep{feroz09} in its python implementation PyMultinest.

\subsection{CHIMERA}
The~\chimera transmission retrieval\footnote{\url{https://github.com/mrline/CHIMERA}} 
tool used in this work is a variant of that described in \cite{line13c,line16,kreidberg15}; \cite{kreidberg18,schlawin18,gharib19,MaiLine19} and \cite{batalha17}. Line-by-line absorption cross-sections come from the pre-tabulated grid described in \cite{freedman14,freedman08}.  As with ~\nemesis and ~\taurex, we convert these line-by-line cross-sections to correlated-K coefficients at a fixed, resolution of 100, with 20 gauss quadrature points (using numpy's gaussquad routine) per wavenumber bin.  As we are not considering multiple scattering in transmission, we need not exploit the resort-rebin procedure here; rather we can simply mix gases by multipyling their transmittances at each g-ordinated bin (per atmospheric layer). 
\section{Forward Model Comparison}
\label{forwards}
The first step of the retrieval comparison was to check that the forward models in each case showed reasonable agreement. We compared output transmission spectra for simple model atmospheres including only a single spectrally active gas, with isothermal temperature profiles. The input bulk properties for the planet were identical in each case, and we tested a range of volume mixing ratios for each spectrally active gas. 

This proved more challenging than expected, and initially we found large discrepancies between the models. Transmission spectra are particularly sensitive to small variations in the calculation of the altitude grid, as the measured quantity is the effective radius of the planet $R_p$+$z$ at the altitude $z$ and pressure $p$ where the atmosphere becomes optically thick. If the model atmosphere is divided into $N$ levels, starting from zero at the bottom, the altitude at grid level $n$ is computed from pressures $p_{n}$ and $p_{n-1}$, altitude $z_{n-1}$ and the scale height $H_{n-1}$, assuming hydrostatic equilibrium:
\begin{equation}
  z_{n} = z_{n-1} + H_{n-1} \mathrm{ln}(p_{n-1}/p_{n})  
\end{equation}
The scale height $H$ at level $n$ is given by
\begin{equation}
    H_n = kT(z_n)/{\mu}g(z_n)
\end{equation}
where $k$ is the Boltzmann constant, $\mu$ is the mean molecular weight of the atmosphere and $T(z_n)$ and $g(z_n)$ are respectively the temperature and local gravitational acceleration at altitude $z$.  

Initially, not all of the models calculated the local gravitational acceleration in the same way, which was found to be the cause of the discrepancy. In the first iteration, CHIMERA kept gravity fixed with altitude, which resulted in deviations of order 10$\%$ in the altitude at the top of the atmosphere. Once this was corrected, smaller differences were found to be caused by factors such as the quoted precision of the gravitational constant $G$ and in the atomic mass unit, which was affecting the calculation of the scale height. Factors affecting the gravitational acceleration $g$ are particularly pernicious because they increase exponentially going up the altitude grid: differences in $g(z_{n-1})$ result in differences in $H_{n-1}$; differences in $H_{n-1}$ result in differences in $z_{n}$, which in turn result in differences in $g(z_{n})$ and so on, leading to a substantial discrepancy in the apparent size of the planet at pressures to which transmission spectroscopy is sensitive. Once these issues were resolved we were able to obtain excellent agreement.

We then moved on to comparing more realistic planet models, including simple clouds and combinations of spectrally active gases (Section~\ref{realforwards}). All of the forward model spectra generated, including input parameters, are made available online\footnote{https://tinyurl.com/y4w7yfzo}. We also include example plotting scripts to facilitate the use of these models in future benchmarking exercises. These scripts, and the retrieval plots presented in this paper, rely on the corner.py package \citep{corner}. 

Most opacity information in each case was taken from the database maintained by ExoMol \citep{tennyson16}. A summary of sources of line data for each molecule is provided in Table~\ref{linedata}. 
\begin{table}
\centering 
\begin{tabular}[c]{r|l|} 
Gas & Source\\
\hline
H$_2$O & \citet{barber06}\tablefootnote{Used by TauREx and NEMESIS}, \citet{Partridge1997}\tablefootnote{Used by CHIMERA}\\
CO$_2$ & \citet{tashkun11,huang14}\\
CO & \citet{roth10,li15}\\
CH$_4$ & \citet{yurchenko14}\\
NH$_3$ & \citet{yurchenko11}\\
\linebreak
\end{tabular}
\caption{List of sources for line data used in forward models. \label{linedata}}
\end{table}

\subsection{Simple Forward Models}
\label{simpleforwards}
Our initial simple forward model test aims to compare the implementation of a model atmosphere with identical properties across the three code suites. This means that we use correlated k tables computed in code-appropriate format from absorption line databases with high completeness; identical gas abundances; and identical temperature profiles. Any deviation between the models is therefore a result of differences in the radiative transfer implementation.

\begin{figure*}[h]
\centering
\includegraphics[width=0.82\textwidth]{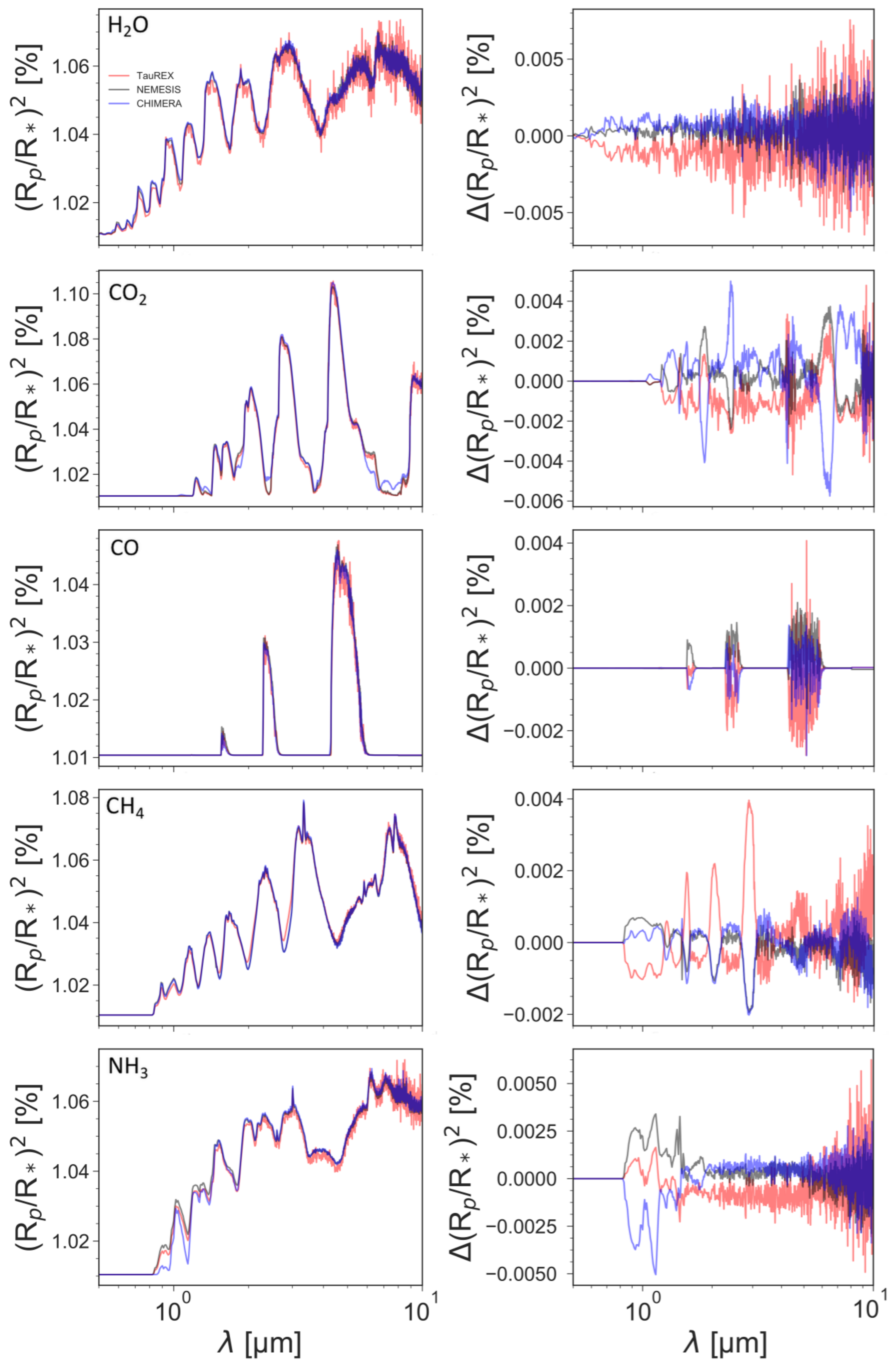}
\caption{Forward model outputs from \taurex, \nemesis~and \chimera~for atmospheres composed of a single spectrally active gas are compared, at temperatures of 1500 K and an abundance of 100 ppmv. Deviations from the average are shown in the right hand column.  \label{forwards_simple}}
\end{figure*}

The first case tested is a 1.0\,M$_\mathrm{J}$, 1.0\,R$_\mathrm{J}$ planet orbiting a 1.0\,R$_{\odot}$ star. We test isothermal profiles of 500\,K, 1000\,K, 1500\,K and 2000\,K. For each of the gases H$_2$O, CO$_2$, CO, CH$_4$ and NH$_3$ we model cases with atmospheres composed of 1\,ppmv, 10\,ppmv, 100\,ppmv, 0.1\,\%, 1\,\%, 10\,\% and 100\,\% of the gas with the remainder a solar composition mixture of H$_2$ and He. We do not include scattering or collision-induced absorption in the initial test. Examples of these simple forward models at 1500\,K and 100 ppmv are shown in  Figure\,\ref{forwards_simple}. Comparisons for other temperatures and abundances are available online.

In general, we find excellent agreement between the forward models, with the only exception being small differences in the wings of the CO and CO$_2$ features, for which~\taurex and ~\nemesis have slightly reduced opacity compared to CHIMERA. In practice, for atmospheres with multiple trace gas components this difference would be unlikely to be noticeable, as we demonstrate in Section~\ref{realforwards}. This exercise established that the three radiative transfer calculations produce comparable results where single spectrally active gases are considered. 

\subsection{More realistic forward models}
\label{realforwards}
Next, we tested more realistic forward models combining spectrally active gases, collision-induced absorption and simple clouds. These models are summarized in Table~\ref{realmods}. Models 0 to 3 have H$_2$-He dominated atmospheres, with a ratio of H$_2$:He of 0.85:0.15. Models 0 and 1 are based on bulk properties for HD\,189733b, whereas models 2--4 represent a warm super-Earth , similar to GJ\,1214b. Clouds are assumed to be optically thick and grey, and are represented by setting the limb transmittances to zero at the specified cloud top pressure (the equivalent of setting the bottom of the atmosphere at this level). We do not explore the effect of including partially transparent or non-grey clouds in this work. Collision-induced absorption due to H$_2$ and He is present in models 0--3. 

\begin{table*}

\centering 
\begin{tabular}[c]{|c|c|c|c|c|c|} 
\hline
Model & Mass, radius, star radius & Temperature (K) & Gases & Clouds & MMW (amu)\\
\hline
0 & 1.162 M$_\mathrm{J}$, 1.138 R$_\mathrm{J}$, 0.781 R$_\odot$ & 1500 & 300 ppmv H$_2$O, 350 ppmv CO & None & 2.3\\
1 & 1.162 M$_\mathrm{J}$, 1.138 R$_\mathrm{J}$, 0.781 R$_\odot$ & 1500 & 300 ppmv H$_2$O, 350 ppmv CO & 10 mbar & 2.3\\
2 & 0.02 M$_\mathrm{J}$, 0.238 R$_\mathrm{J}$, 0.216 R$_\odot$ & 400 & 800 ppmv H$_2$O, 400 ppmv CH$_4$, 100 ppmv NH$_3$ & None & 2.3\\
3 & 0.02 M$_\mathrm{J}$, 0.238 R$_\mathrm{J}$, 0.216 R$_\odot$ & 400 & 800 ppmv H$_2$O, 400 ppmv CH$_4$, 100 ppmv NH$_3$ & 10 mbar & 2.3\\
4 & 0.02 M$_\mathrm{J}$, 0.238 R$_\mathrm{J}$, 0.216 R$_\odot$ & 400 & 0.43 H$_2$O, 0.25 CO$_2$, 0.14 CH$_4$, 0.13 N$_2^{*}$ & None & 24.7\\
\hline
\end{tabular}
\caption{Details of the five `realistic' model planets. $^*$Note that we do not include collision induced absorption for N$_2$. \label{realmods}}
\end{table*}

\begin{figure*}
\centering
\includegraphics[width=0.85\textwidth]{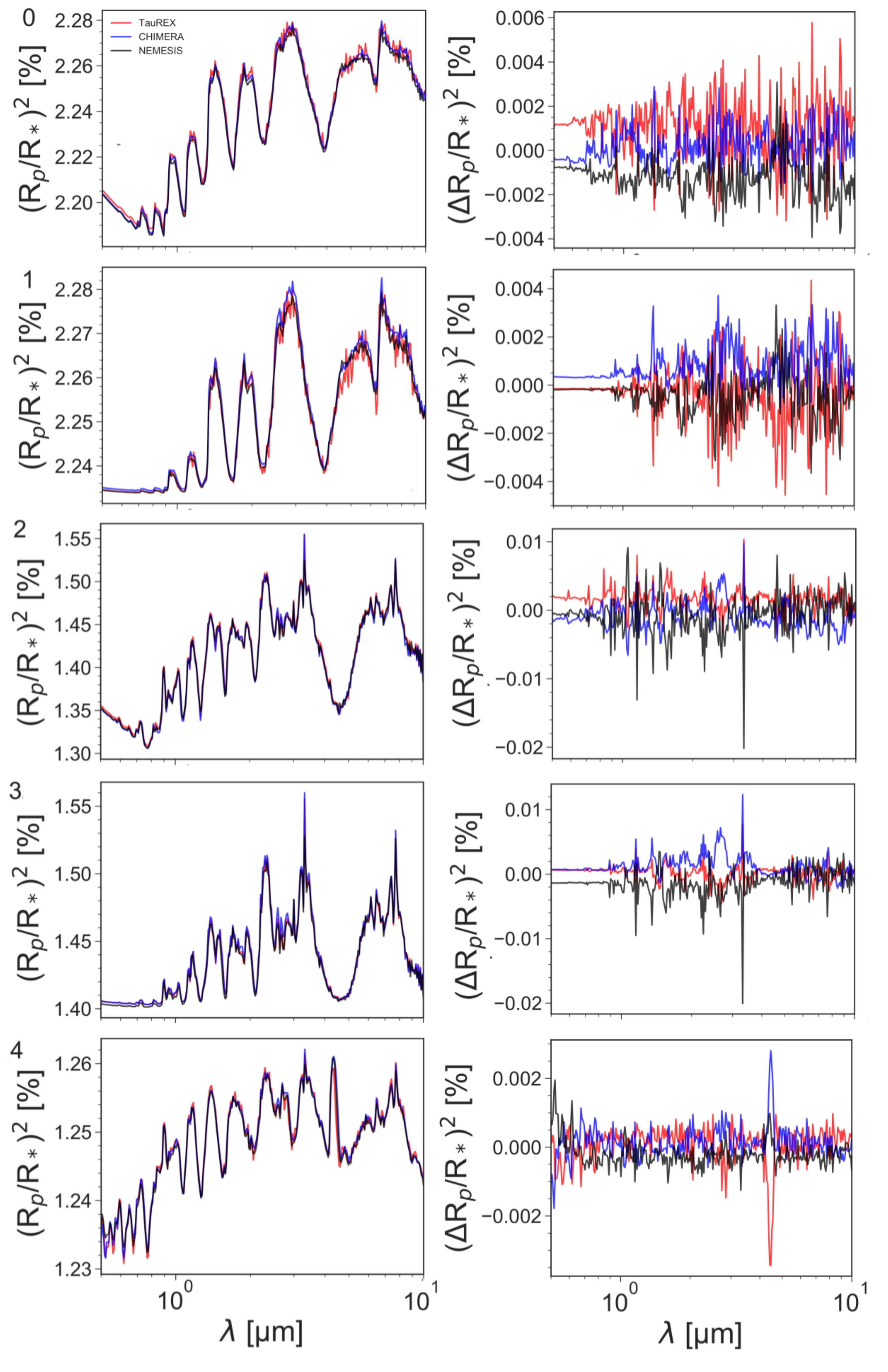}
\caption{Forward model outputs from \taurex, \nemesis~and \chimera~for more realistic planet models 0 - 4. These represent cloud-free and cloudy hot Jupiter atmospheres. Residuals from the average of the three spectra are shown in the right hand column. \label{hotjupreal}}
\end{figure*}

We also find agreement with these more complex models (Figure~\ref{hotjupreal}). The interplay between collision-induced continuum absorption, clouds and molecular absorption is generally reproducible between different atmospheric models, at the level of a few 10s of ppm for the most part. 

Exceptions to this occur for the centre of the $\nu$3 band of methane at 3.3\,$\upmu$m for the cloud-free super Earth case. The core of the CH$_4$ feature is both narrow and high amplitude, and as such this type of feature is most likely to be affected by small differences in forward model calculation. However, as this difference occurs for only a single spectral point, the impact in a low-resolution retrieval scenario would be small as the feature would be smoothed. The current state-of-the-art achievable with \textit{HST} will be unaffected, but the effects may need to be considered with \textit{JWST} quality spectra. This will be tested in Section~\ref{retrievals}.

\begin{figure}
\centering
\includegraphics[width=0.95\columnwidth]{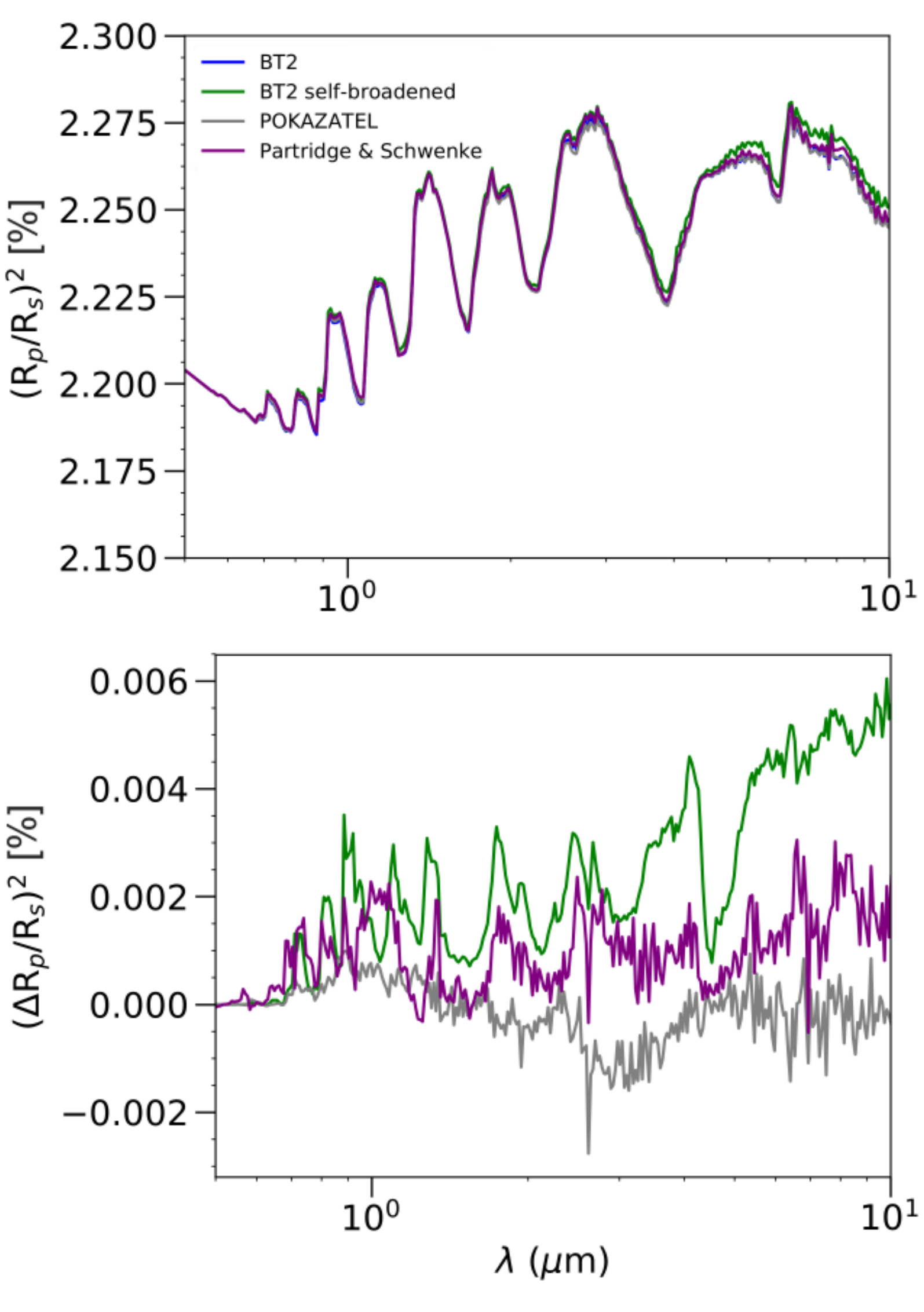}
\caption{Forward models generated using CHIMERA, containing different H$_2$O linelists and broadening assumptions, are presented here. A difference plot against the H2-He broadened BT2 linelist is also shown. The assumption of self-broadening vs H2-He broadening has the most significant difference, whereas the other differences are $<$20 ppm.  \label{linedata_test}}
\end{figure}

Differences between the models are most likely to be attributable to small deviations in the gas opacity data used, either in the linelist itself or the tabulation thereof. To demonstrate the sensitivity of the spectra to differences in both the source of opacity data, and the treatment of line broadening, we present four Planet 0 spectra generated by CHIMERA in Figure~\ref{linedata_test}. As well as the BT2 linelist\citep{barber06} with H2-He broadening, we include the same linelist with the assumption of self-broadening, which is $\sim$ 7$\times$ larger than the H2-He broadening coefficient \citep{gharib19}; the new POKAZATEL linelist \citep{pokazatel}, which was produced after the bulk of the work for this paper was performed and is considered to be the new state-of-the-art; and a theoretical linelist produced using the potential energy surfaces derived by \cite{Partridge1997}, generated as described in \cite{freedman14}.  The POKAZATEL, BT2, \& BT2 Self-Broadened were generated following the exact methods described in \cite{gharib19}. The cross-sections computed using the BT2 with H2-He broadening, BT2 with self-broadening, and POKOZATEL (also with H2-He broadening) assume a 100cm$^{-1}$ Voight line-wing cutoff whereas those computed from \cite{Partridge1997} assume a 25cm$^{-1}$ cutoff.  Comparing POKOZATEL to BT2 H2-He broadened cross-sections illustrates  the influence of {\it line-list choice}, all else (broadening species, Voight-wing cutoff, sampling resolution) being equal. Comparing the BT2 self-broadened spectrum to BT2 H2-He broadened spectrum illustrates the influence in {\it broadening species}, all else being equal (see also \cite{gharib19} for a more detailed comparison). Finally, comparison of the latter 3 scenarios to the \cite{Partridge1997} scenario illustrates simultaneously the differences between line-list choice and Voigt wing cutoff. By far the most significant difference is caused by using H$_2$O self broadening rather than H$_2$-He broadening, indicating that for linelists with sufficiently high completeness the exact list used is a secondary consideration relative to correct treatment of line broadening.

This section demonstrates the challenge of producing identical outputs from different model frameworks, even when attempts have been made to align the inputs and processes used. Whilst we achieve a high level of agreement, differences still remain at the level of several 10s of ppm. We explore the effect of this on retrieval scenarios in the next section.

\section{Retrieval Comparison}
We take the five model planets introduced in Section~\ref{realforwards} and bin the spectra down to a resolution of R=100 over the wavelength range of 0.5--10 $\upmu$m. This wavelength range covers the majority of the \textit{JWST} range and also the spectral range of the recently-selected \textit{ARIEL} spacecraft, which will perform a dedicated atmospheric census for transiting exoplanets \citep{tinetti18}. These spectra are cross-retrieved between the three algorithms to assess whether spectra generated with one model can be accurately retrieved using the others. We test error bars at 30, 60 and 100 ppm. We do not randomise the spectral points due to noise, or perform self-retrievals, as in this case we are testing the retrieval algorithms against each other rather than the performance of the individual retrievals on data, and outliers within a noise draw might hinder our ability to make a meaningful comparison (e.g., \cite{feng2018}). 

Retrieved quantities in each case include the abundances of the gases known to be included in the model; the isothermal temperature; the cloud top pressure; and the radius of the planet at 10 bar. The radius retrieval is required when analysing primary transit spectra because the pressure level to which the radius derived from the white light transit corresponds is not known. This is especially true in the case of a cloudy atmosphere, where the white light transit radius may actually correspond to the cloud top rather than the level at which a cloud-free atmosphere would become opaque. We choose to retrieve the radius at 10 bar because we consider that at this pressure all atmospheres are opaque in transit geometry at visible and infrared wavelengths, though in general the reference pressure is arbitrary and rather inconsequential since the radius is a free parameter.

We then use each of the retrieval codes in turn to model synthetic observations generated by the other two, for each example planet. The results are compared with the known input values and with each other. All retrieval results with full posterior distributions are available via the online repository, including the MultiNest output and example plotting routines; we present a subset here to illustrate our main findings.   

\begin{figure*}
\centering
\includegraphics[width=0.8\textwidth]{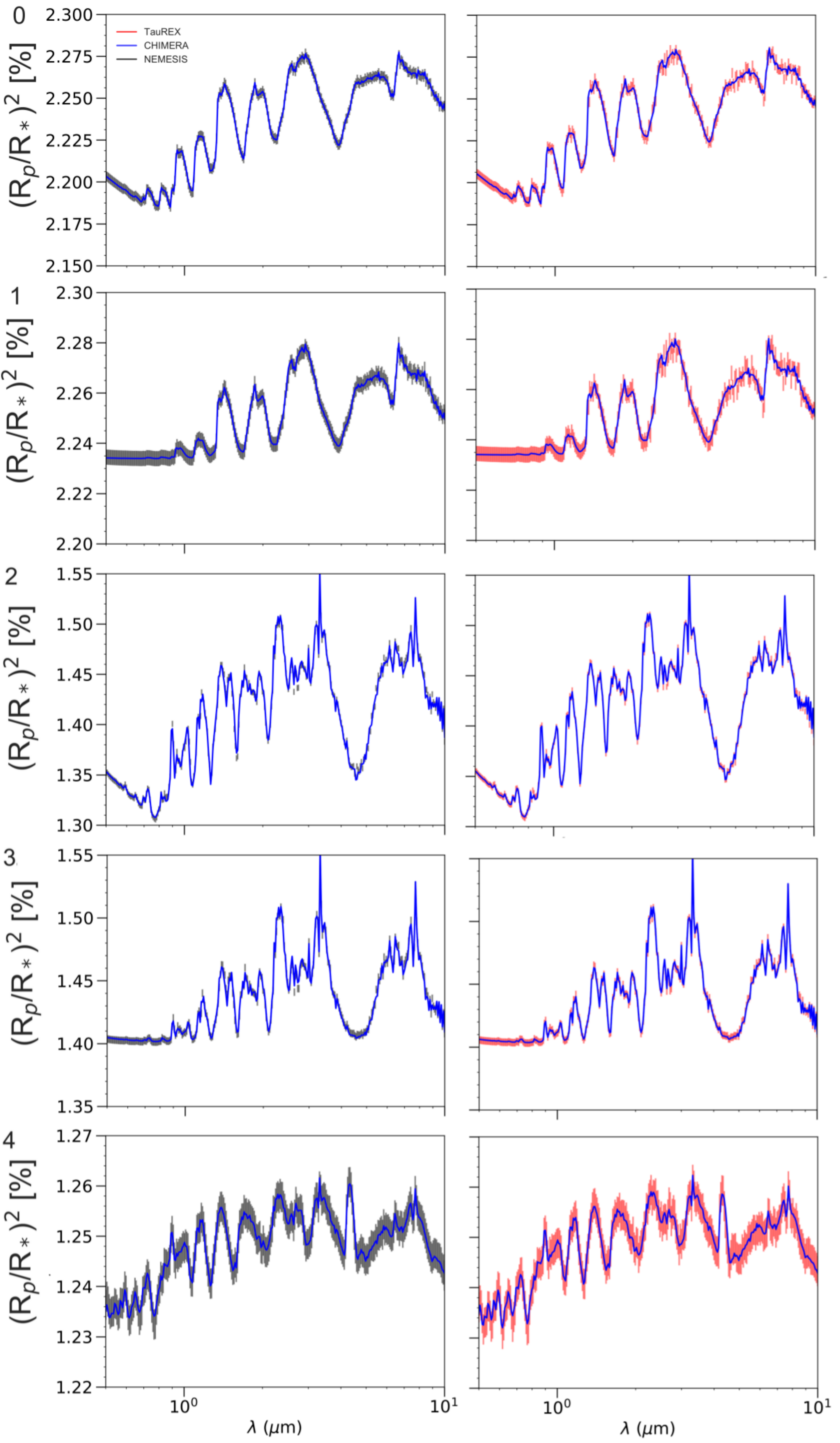}
\caption{We show the best fit retrieved spectra from CHIMERA (blue solid curve) on synthetic observations generated using NEMESIS (black ticks, left column) and TauREX (red ticks, right column) for each of our `realistic' model planets, assuming a 30 ppm error envelope. As expected given the results of our forward model comparison, the quality of the fit is very good, typically falling within the simulated error envelope.  \label{chimera_bestfits}}
\end{figure*}

\begin{figure*}
\centering
\includegraphics[width=0.8\textwidth]{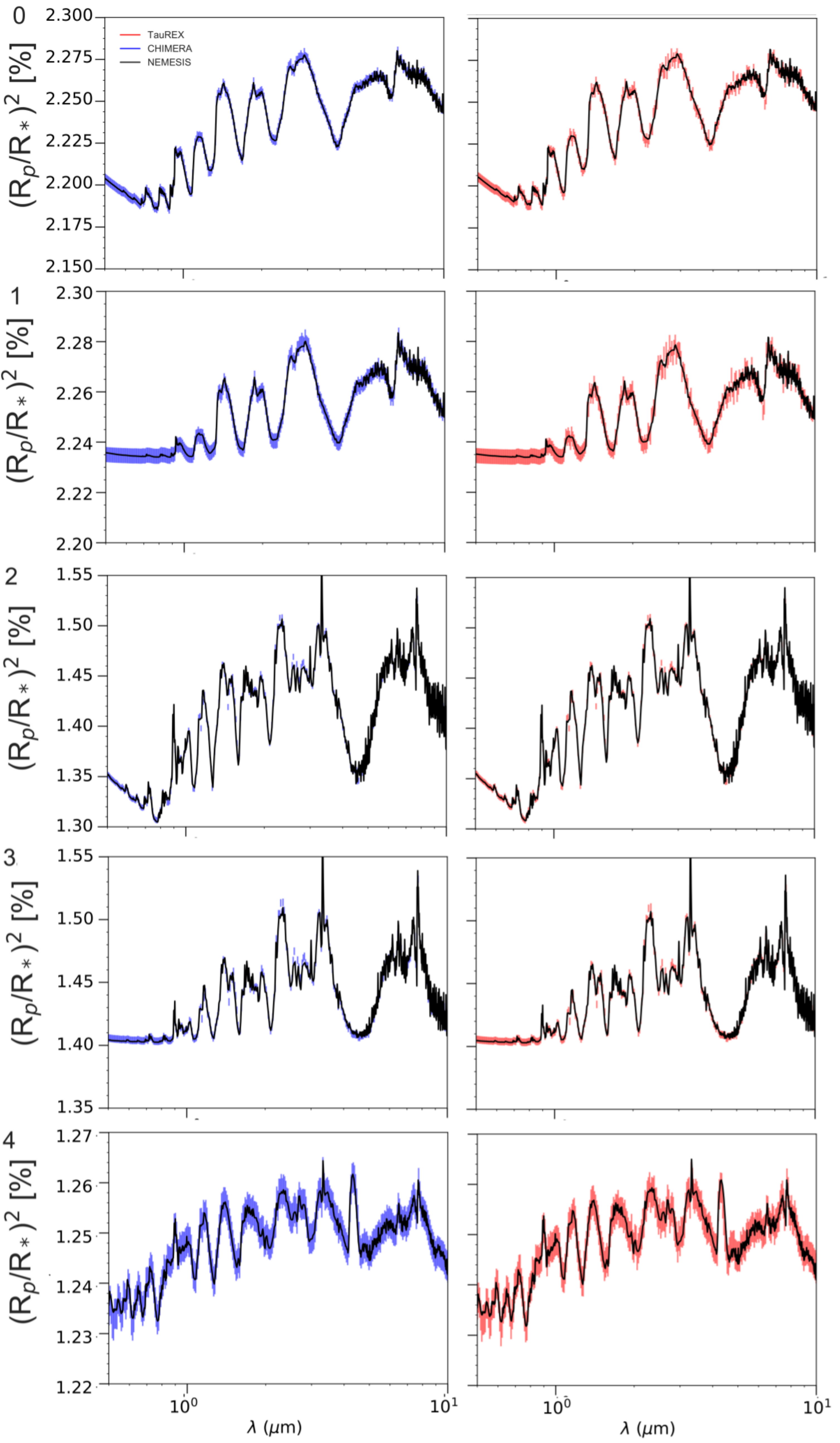}
\caption{As Figure~\ref{chimera_bestfits} for ~\nemesis (black solid curve) on ~\chimera (blue error envelope, left column) and ~\taurex (red error envelope, right column). \label{nemesis_bestfits}}
\end{figure*}

\begin{figure*}
\centering
\includegraphics[width=0.8\textwidth]{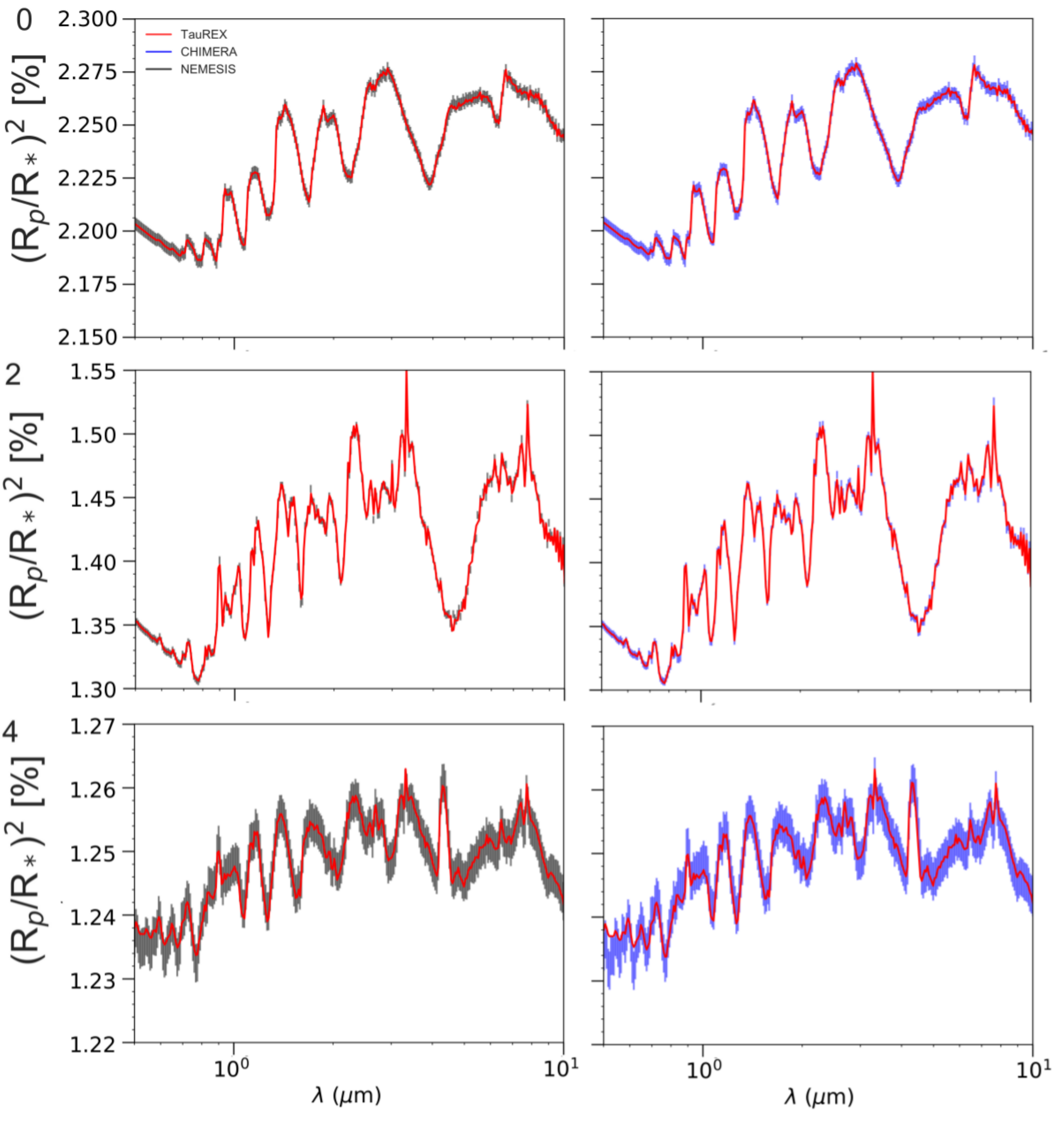}
\caption{As Figure~\ref{chimera_bestfits} for ~\taurex (red solid curve) on ~\nemesis (black error envelope, left column) and ~\chimera (blue error envelope, right column). We include only the cloud-free fits here, because ~\taurex recovers multiple probability maxima for the cloudy scenarios with a 30 ppm error envelope.\label{taurex_bestfits}}
\end{figure*}

\begin{figure*}
\centering
\includegraphics[width=0.8\textwidth]{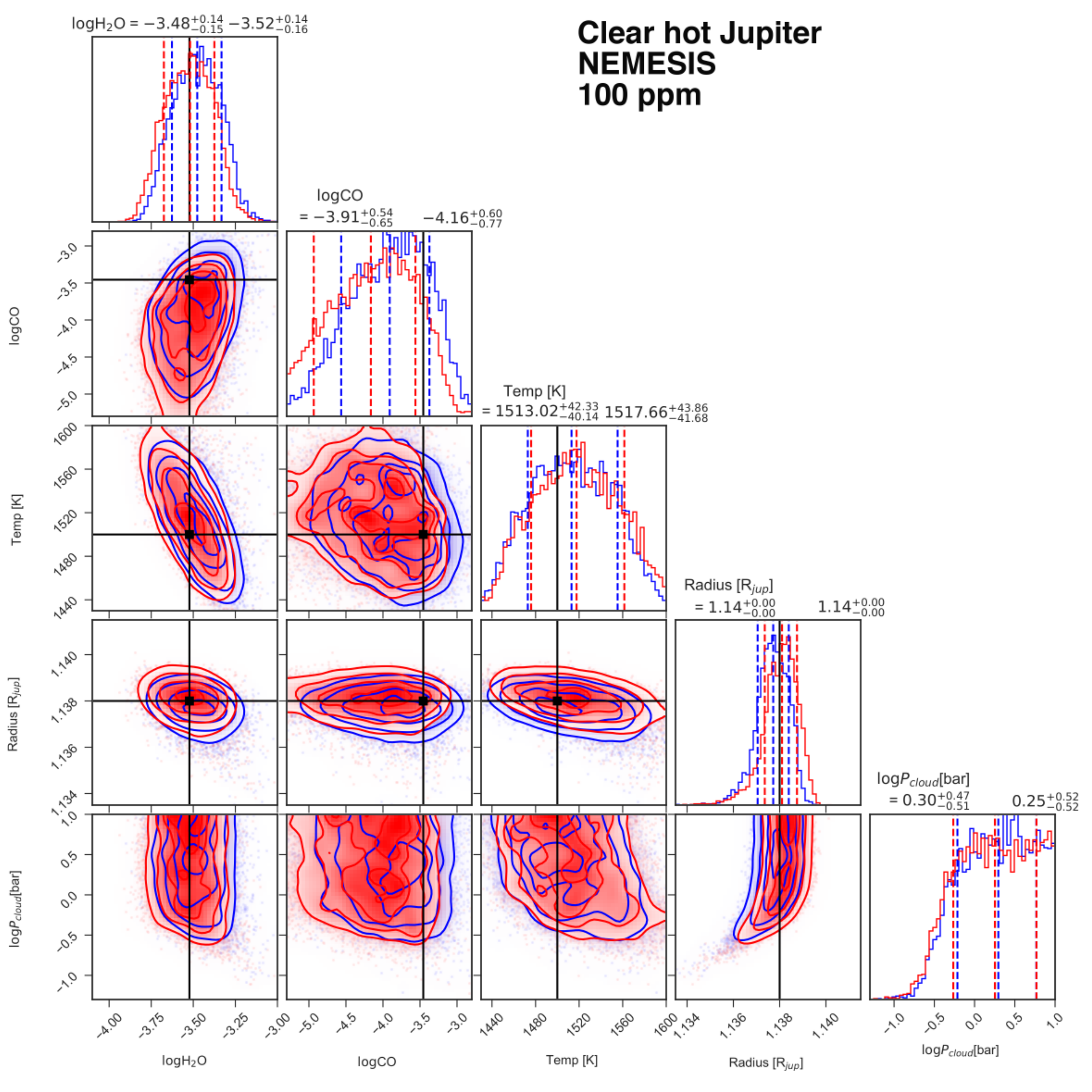}
\caption{Cross-retrievals for ~\nemesis on ~\taurex (red) and ~\chimera (blue) for Planet 0, a cloud-free hot Jupiter. The black lines denote the true input values.  \label{planet0_nemesis_100}}
\end{figure*}

\begin{figure*}
\centering
\includegraphics[width=0.8\textwidth]{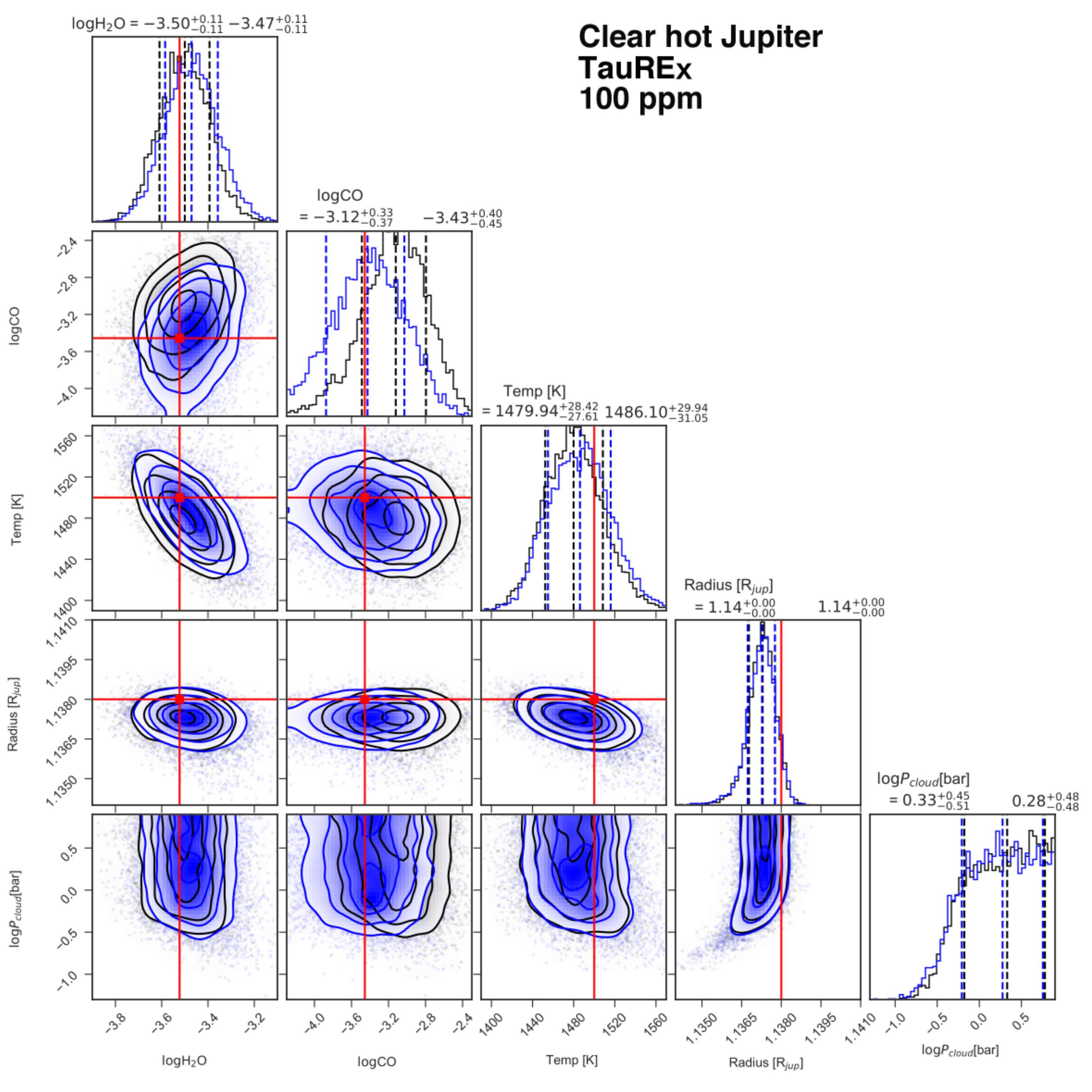}
\caption{As Figure~\ref{planet0_nemesis_100} for ~\taurex on ~\nemesis (black) and ~\chimera (blue). The red lines denote the true input values.\label{planet0_taurex_100}}
\end{figure*}

\begin{figure*}
\centering
\includegraphics[width=0.8\textwidth]{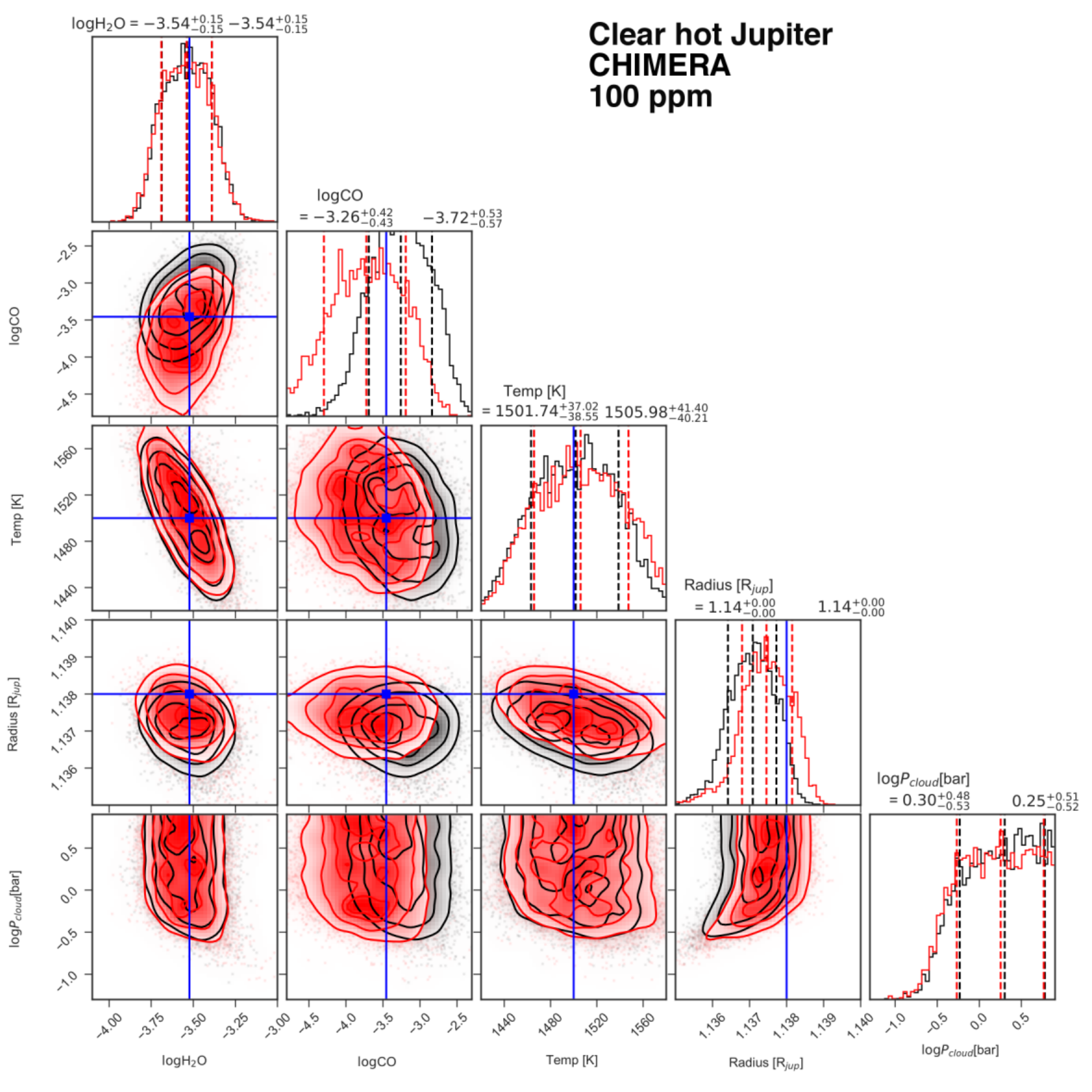}
\caption{As Figure~\ref{planet0_nemesis_100} for ~\chimera on ~\nemesis (black) and ~\taurex (red). The blue lines denote the true input values.\label{planet0_chimera_100}}
\end{figure*}

\subsection{Retrieval Comparison Results}
\label{retrievals}
We perform cross-retrievals for each code on synthetic observations produced by both of the other two codes, assuming either 30, 60 or 100 ppm error envelopes. To demonstrate the quality of the fit we achieve, in Figures~\ref{chimera_bestfits}--\ref{taurex_bestfits} we show the retrieved spectra against the synthetic observations for each code and planet, assuming a 30 ppm error envelope. For ~\taurex we show only the cloud-free cases, because ~\taurex recovers multiple, separate probability maxima for these cases, meaning that a single best-fit spectrum would not be representative. It is clear from these figures that, unsurprisingly given the extensive benchmarking of the forward models, all retrievals produce best fit spectra that match the synthetic observation well.

We discuss cross-retrievals for each planet below. Planet 0, the cloud-free hot Jupiter case, represents the 'low hanging fruit' of transiting exoplanet targets. Hot Jupiters have large scale heights, and those with clear atmospheres are expected to have the clearest molecular absorption features within their transit spectra.

First, we present the results of all six cross-retrievals assuming a 100 ppm error envelope (\nemesis on \taurex, \nemesis on \chimera, \taurex on \nemesis, \taurex on \chimera, \chimera on \nemesis and \chimera on TauREx). The retrieved posteriors are shown for the NEMESIS, \taurex and \chimera retrievals in Figures~\ref{planet0_nemesis_100} to~\ref{planet0_chimera_100} respectively. The colour scheme from above is maintained here, with \nemesis shown in black, \taurex in red and \chimera in blue. 

In Figures~\ref{planet0_nemesis_100} -- \ref{planet0_chimera_100} it is clear that in all cases H$_2$O, Temperature and cloud top pressure (in this case, high cloud top pressure with a lower limit correctly indicates an absence of cloud) are correctly retrieved to within 1$\sigma$. The \nemesis on \taurex retrieval slightly underestimates the CO abundance, and both \taurex retrievals and the \chimera on \nemesis retrieval slightly underestimate the radius. In general, H$_2$O and temperature are most reliably retrieved, and the retrieved values are almost the same regardless of the retrieval model and input models used. 

We expect H$_2$O to be more reliably constrained than CO as there are multiple strong H$_2$O features in the spectrum, whereas CO only has a single feature at 4.5\,$\upmu$m strong enough to be detected above the H$_2$O (see Figure~\ref{forwards_simple}). The slight underestimations of the radius are likely to be due to the inclusion of the grey cloud deck in the retrieval model when no cloud is present in the input spectrum; a very deep, grey cloud is indistinguishable from a slight offset in the radius. 

\begin{figure*}
\centering
\includegraphics[width=0.8\textwidth]{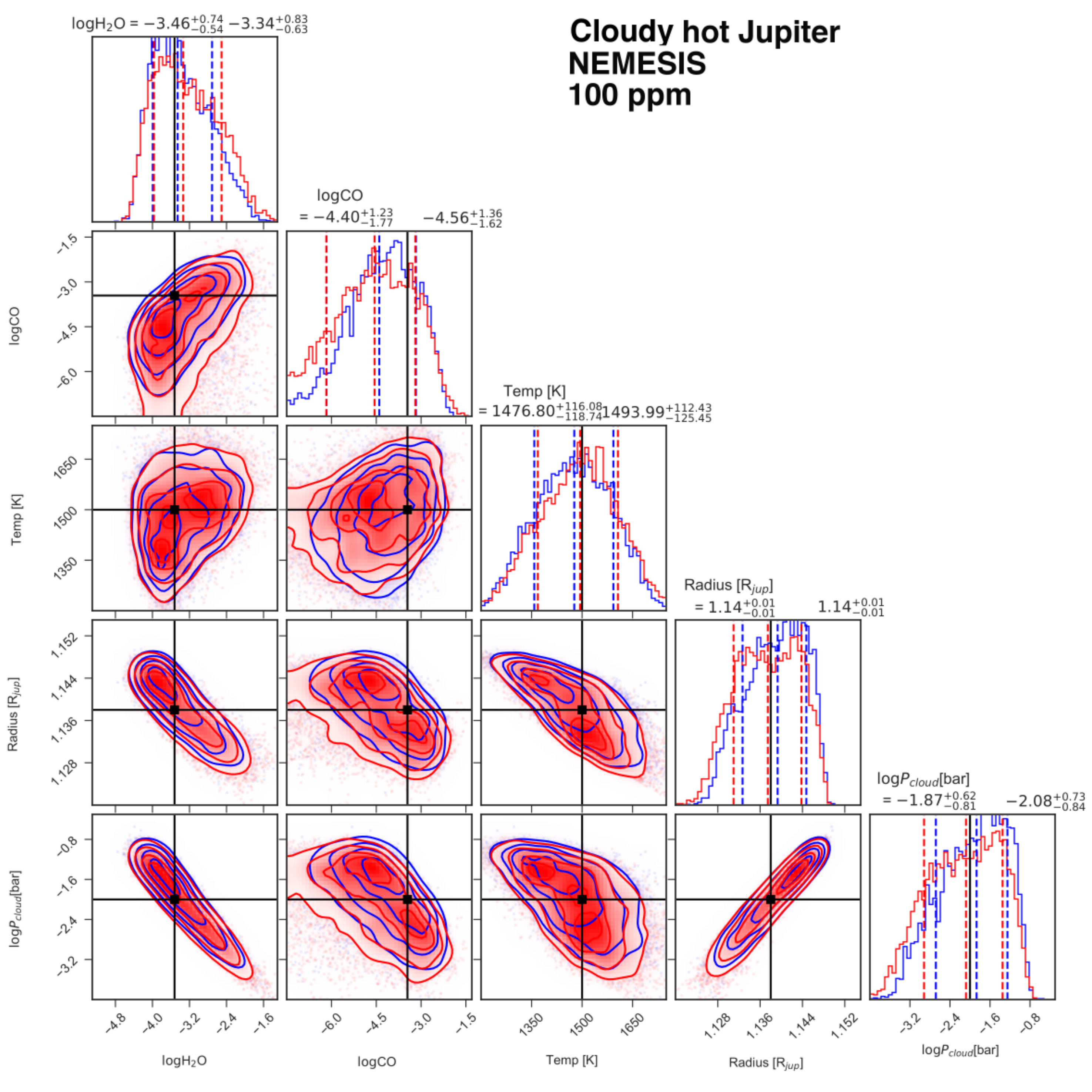}
\caption{Cross-retrievals for \nemesis on \taurex (red) and \chimera (blue) for Planet 1, a cloudy hot Jupiter. The black lines denote the true input values.  \label{planet1_nemesis_100}}
\end{figure*}

\begin{figure*}
\centering
\includegraphics[width=0.8\textwidth]{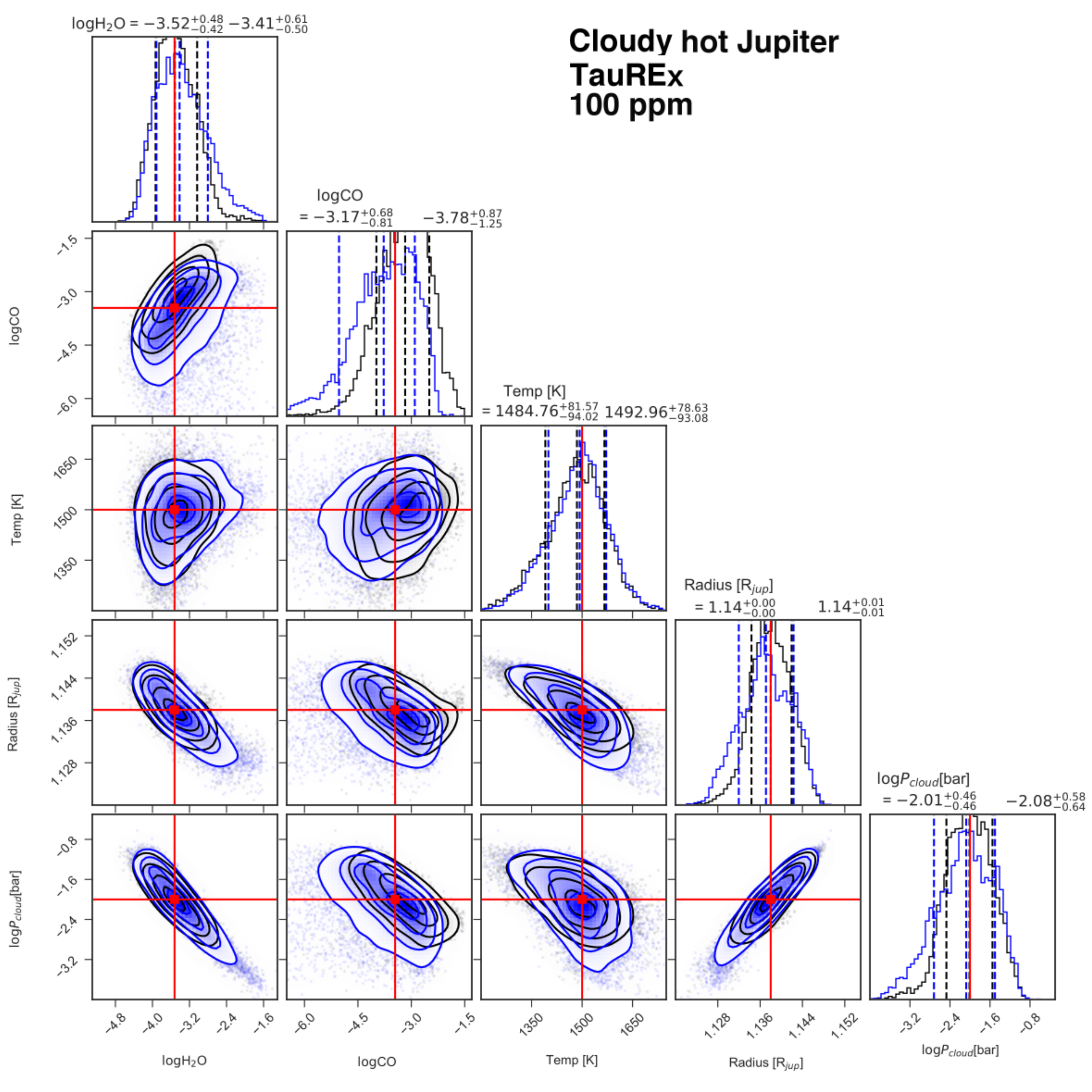}
\caption{As Figure~\ref{planet1_nemesis_100} for \taurex on \nemesis (black) and ~\chimera (blue). The red lines denote the true input values.\label{planet1_taurex_100}}
\end{figure*}

\begin{figure*}
\centering
\includegraphics[width=0.8\textwidth]{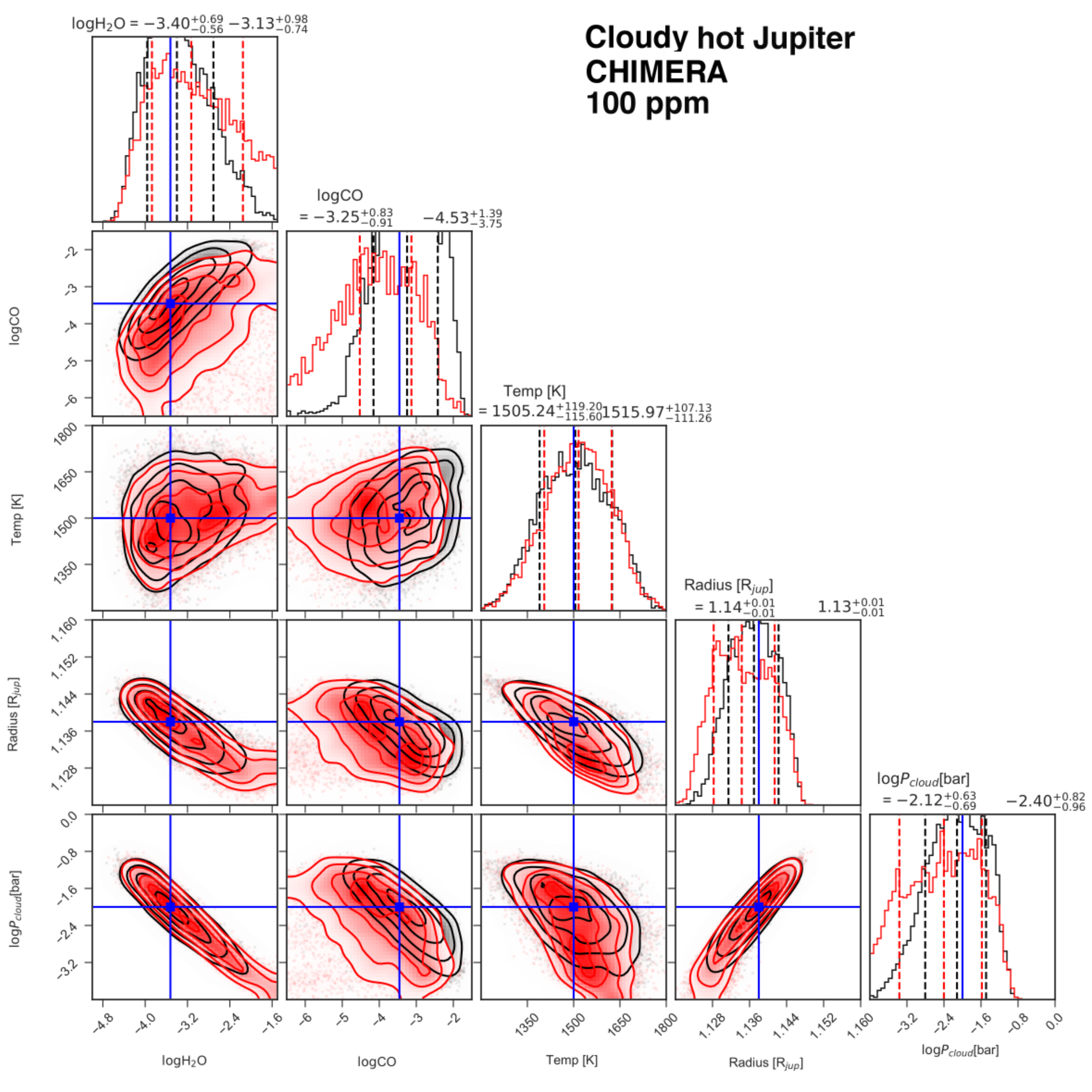}
\caption{As Figure~\ref{planet1_nemesis_100} for \chimera on \nemesis (black) and \taurex (red). The blue lines denote the true input values.\label{planet1_chimera_100}}
\end{figure*}

Comparing retrievals for Planet 0 and Planet 1 is instructive. The two cases are identical except for the fact that Planet 1 contains an opaque, grey cloud deck at 10\,mbar. The same set of 6 retrievals with an error bar of 100 ppm is presented in Figures~\ref{planet1_nemesis_100}--~\ref{planet1_chimera_100}. The posteriors for the cloudy case are much broader, because adding cloud has the effect of flattening the molecular features, thus increasing the degeneracy of solutions. However, in all cases the retrieved quantities are correct to within 1\,$\sigma$; despite the additional degeneracy and decreased precision, the solution accuracy is actually improved for the cloudy case, because we are correctly fitting a cloudy model to a cloudy spectrum. This provides an argument for testing a cloud-free retrieval scenario in the case that initial retrieval results demonstrate a lack of evidence for clouds; indeed, following Occam's Razor, this applies to any parameter when there is little evidence that it is required to fit the spectrum. 

All retrievals for 60\,ppm and 30\,ppm error bars for Planets 0 and 1 are available in our online repository. 

\begin{figure*}
\centering
\includegraphics[width=1.0\textwidth]{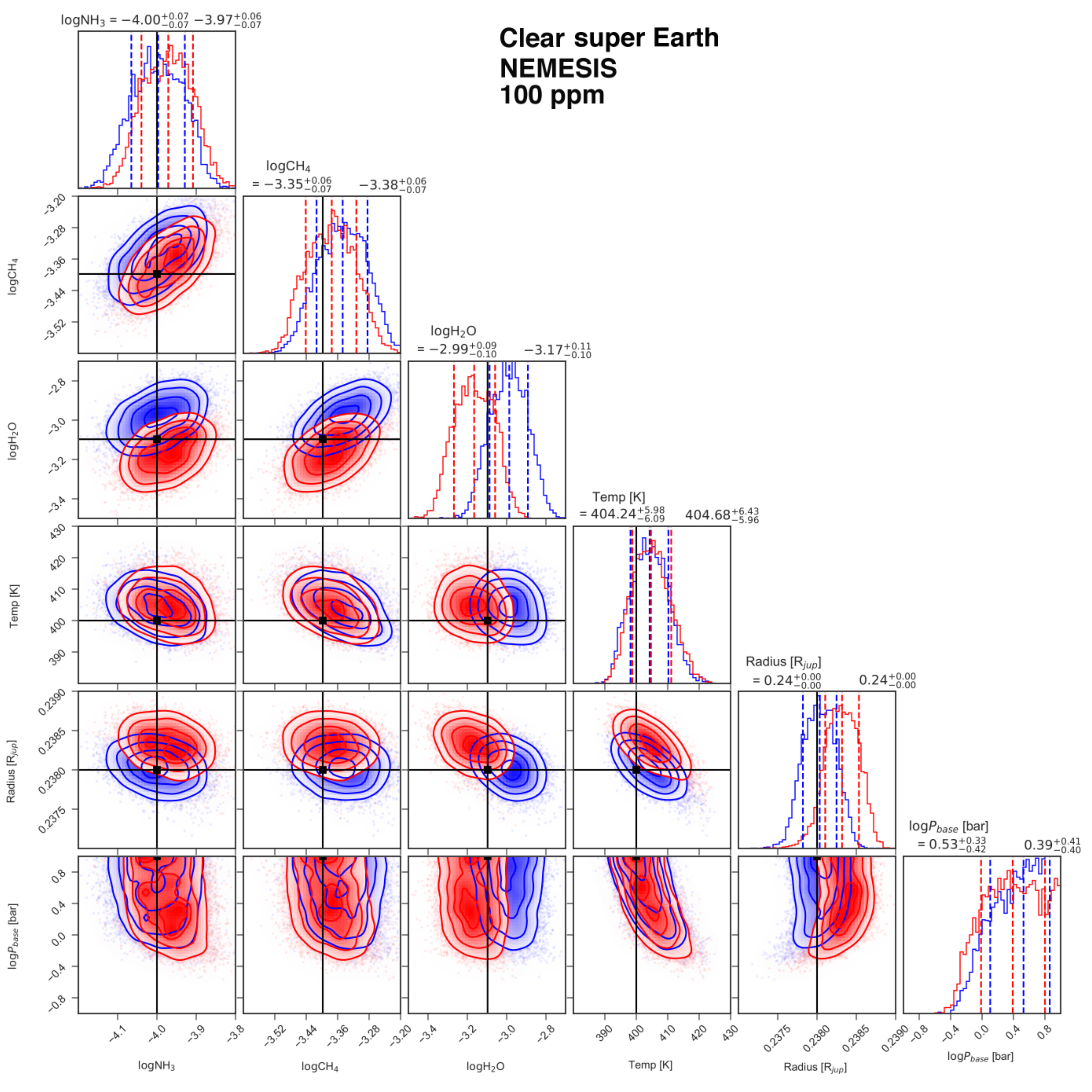}
\caption{Retrievals of Planet 3, 100 ppm, for \nemesis on \chimera (blue) and \taurex (red). The black lines denote the true input values.\label{planet2_nemesis_100}}
\end{figure*}

We now consider the super Earth retrievals (planets 2--4). Super Earths are likely to be more challenging targets; smaller planets have smaller transit depths and are more likely to have small scale heights, as they may have high-mean-molecular-weight atmospheres. Figure~\ref{planet2_nemesis_100} is an example of a cloud-free super Earth retrieval with NEMESIS. As with the cloud-free hot Jupiter case, retrievals are consistent with each other and with the correct solution for a 100\,ppm error bar, and the precision improves (although the accuracy decreases) as the error bar reduces. 

\begin{figure*}
\centering
\includegraphics[width=1.0\textwidth]{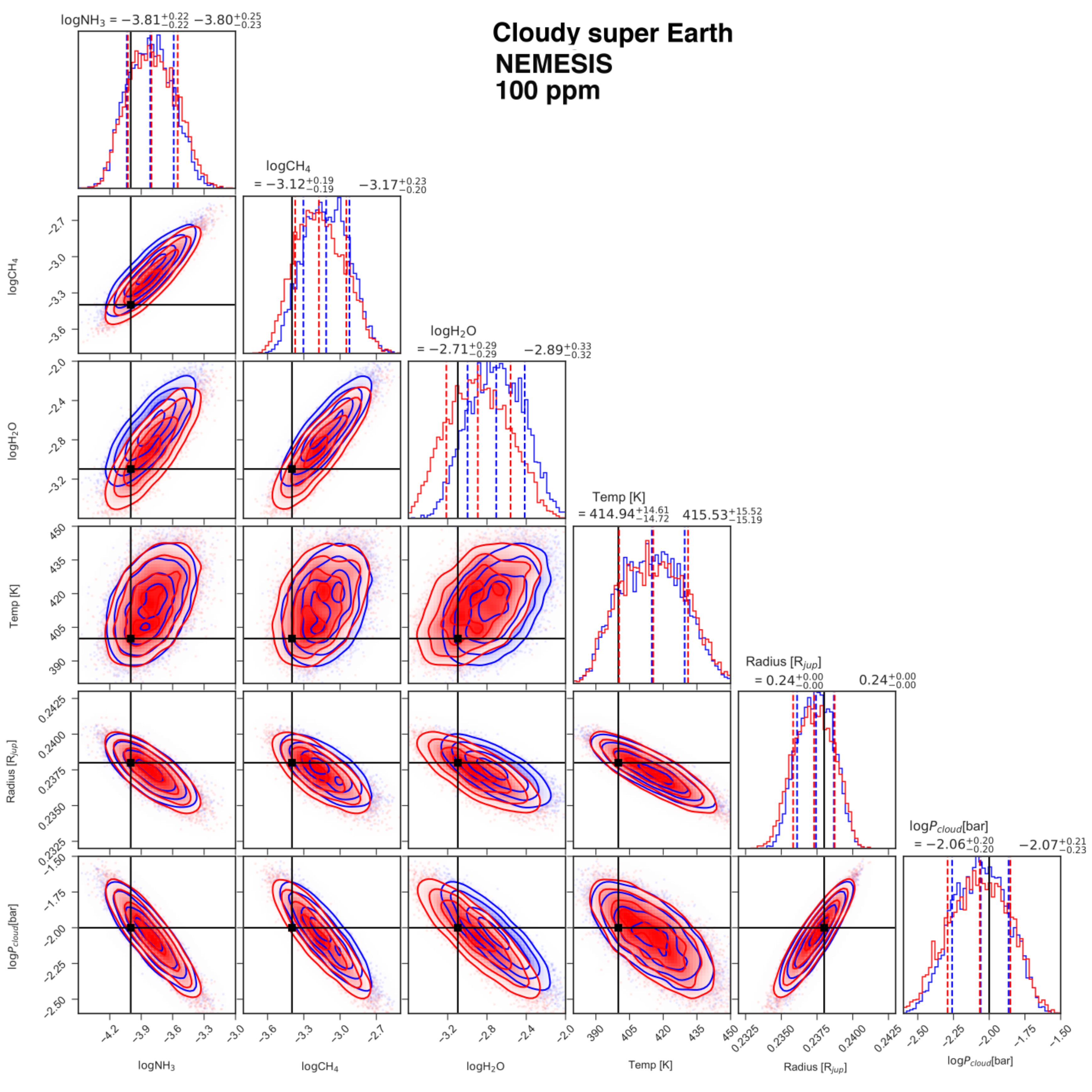}
\caption{Retrievals of Planet 3, 100 ppm, for \nemesis on \chimera (blue) and \taurex (red). The black lines denote the true input values.\label{planet3_nemesis_100}}
\end{figure*}

\begin{figure*}
\centering
\includegraphics[width=1.0\textwidth]{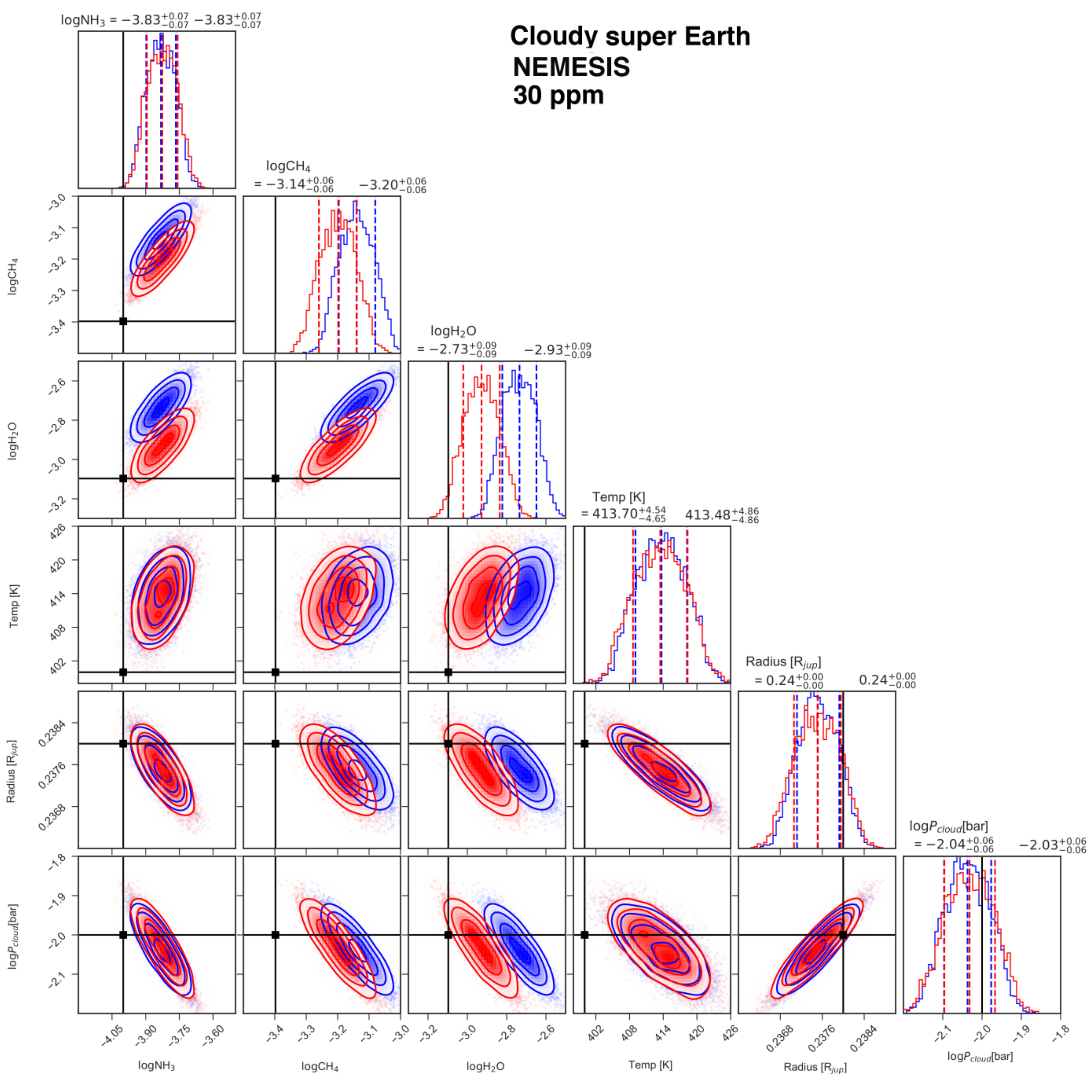}
\caption{As Figure~\ref{planet3_nemesis_100}, but for a 30 ppm error envelope.\label{planet3_nemesis_30}}
\end{figure*}

Here, we present cross retrievals for Planet 3, the cloudy super Earth case. We compare the results for 100 ppm (Figure~\ref{planet3_nemesis_100}) and 30 ppm (Figure~\ref{planet3_nemesis_30}) error bars. There are several points of interest to note. 

At 100 ppm, the retrievals on~\taurex and~\chimera overlap almost exactly for all variables. The exception is for the H$_2$O abundance, which is slightly over-retrieved against the input value for the~\chimera spectrum but not the~\taurex spectrum. CH$_4$ is slightly over-retrieved for both cases. However, the agreement for the two synthetic spectra is extremely good. 

Differences emerge for the case with 30 ppm error bars. In this instance, the retrievals no longer perfectly overlap (although they are still consistent with each other within the 1\,$\sigma$ limits.) The temperature and gas abundances are no longer correctly retrieved to within 1\,$\sigma$, although the values are still very close to the true ones. 

There are several possible origins for these differences. As discussed in Section~\ref{realforwards}, for the super Earth models the centres of the CH$_4$ bands have different amplitudes for~\nemesis vs~\taurex and~\chimera. Since the information about temperature comes from the amplitudes of all molecular features, this discrepancy could affect temperature retrievals as well as the CH$_4$ abundance. There is also a very small baseline offset present between the models in the optical region where there is no molecular absorption, which can be attributed to differences in the way that an opaque grey cloud deck is modelled. Both of these could be responsible for the differences between the retrieved and true values, and the different results for the two forward models. 

We note here that, because \nemesis uses the Importance Nested Sampling option within MultiNest\citep{feroz13}, it is unable to separate multiple minima in multi-modal distributions. We can see this effect by comparing the 30\,ppm \nemesis retrieval with the \taurex retrieval for planet 3, which clearly finds two minima for each retrieval; in the case of the retrieval on ~\nemesis, the minima have roughly equal probability.

\begin{figure*}
\centering
\includegraphics[width=1.0\textwidth]{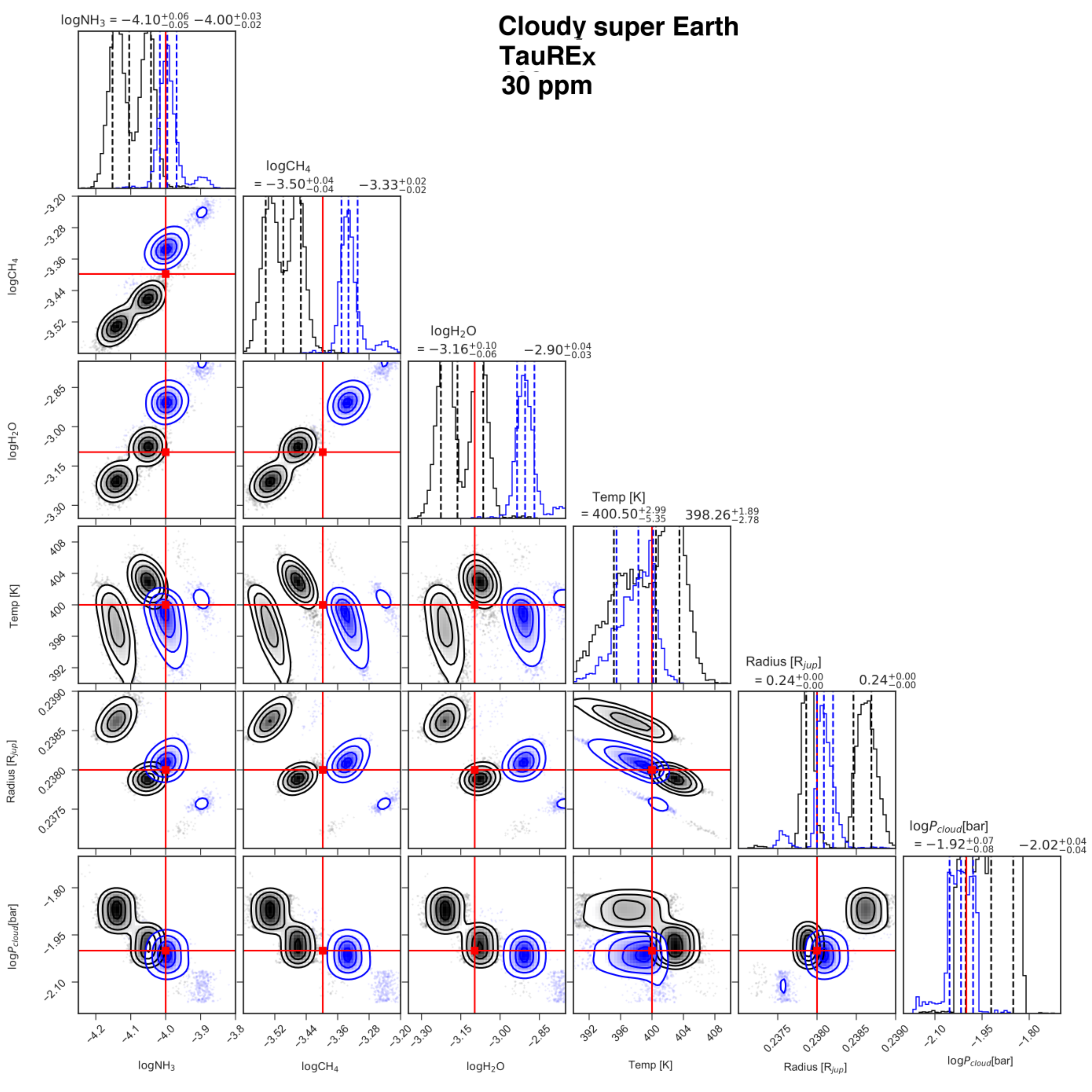}
\caption{As Figure~\ref{planet3_nemesis_30}, but for \taurex retrievals on \nemesis and \chimera spectra. Retrievals on \nemesis are shown in black and on \chimera in blue. Note that multiple minima are present for both retrievals. \label{planet3_taurex_30}}
\end{figure*}

Finally, we examine planet 4 - the high-mean-molecular-weight super Earth. Planets of this size pose a challenge for retrievals, as the bulk composition of the atmosphere and therefore the mean molecular weight is not known \textit{a priori}. This can in theory be calculated based on the retrieved abundances of spectrally active gases, but is challenging in cases where there is still a substantial fraction of a non-spectrally active gas present. Whilst in practice N$_2$ does have some infrared spectral features, in this case we include N$_2$ without any spectroscopic information. The recovery of the correct N$_2$ abundance is therefore entirely dependent on retrieving all spectrally active gases correctly, and thence calculating the mean molecular weight. In this case, we have made the assumption that the atmosphere is high mean molecular weight and therefore the non-spectrally active part of the atmosphere is N$_2$ rather than a H$_2$-He mix; for lower density planets, where this may be ambiguous, the problem will be more difficult to solve. 

\begin{figure*}
\centering
\includegraphics[width=1.0\textwidth]{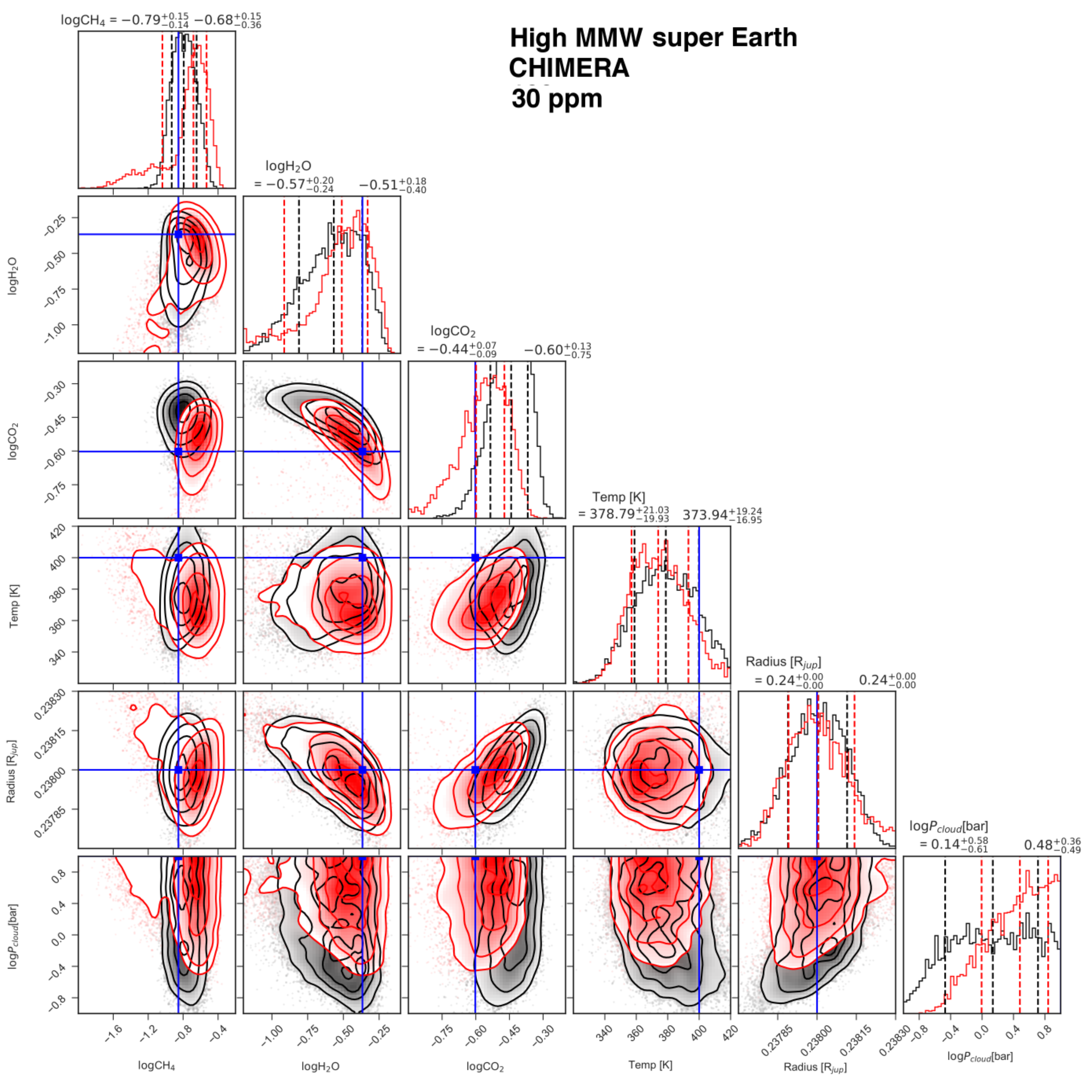}
\caption{Retrieval using \chimera on spectra from \taurex (red) and \nemesis (black) for Planet 4, the high-mean-molecular-weight super Earth. \label{planet4_chimera_30}}
\end{figure*}

\begin{figure*}
\centering
\includegraphics[width=1.0\textwidth]{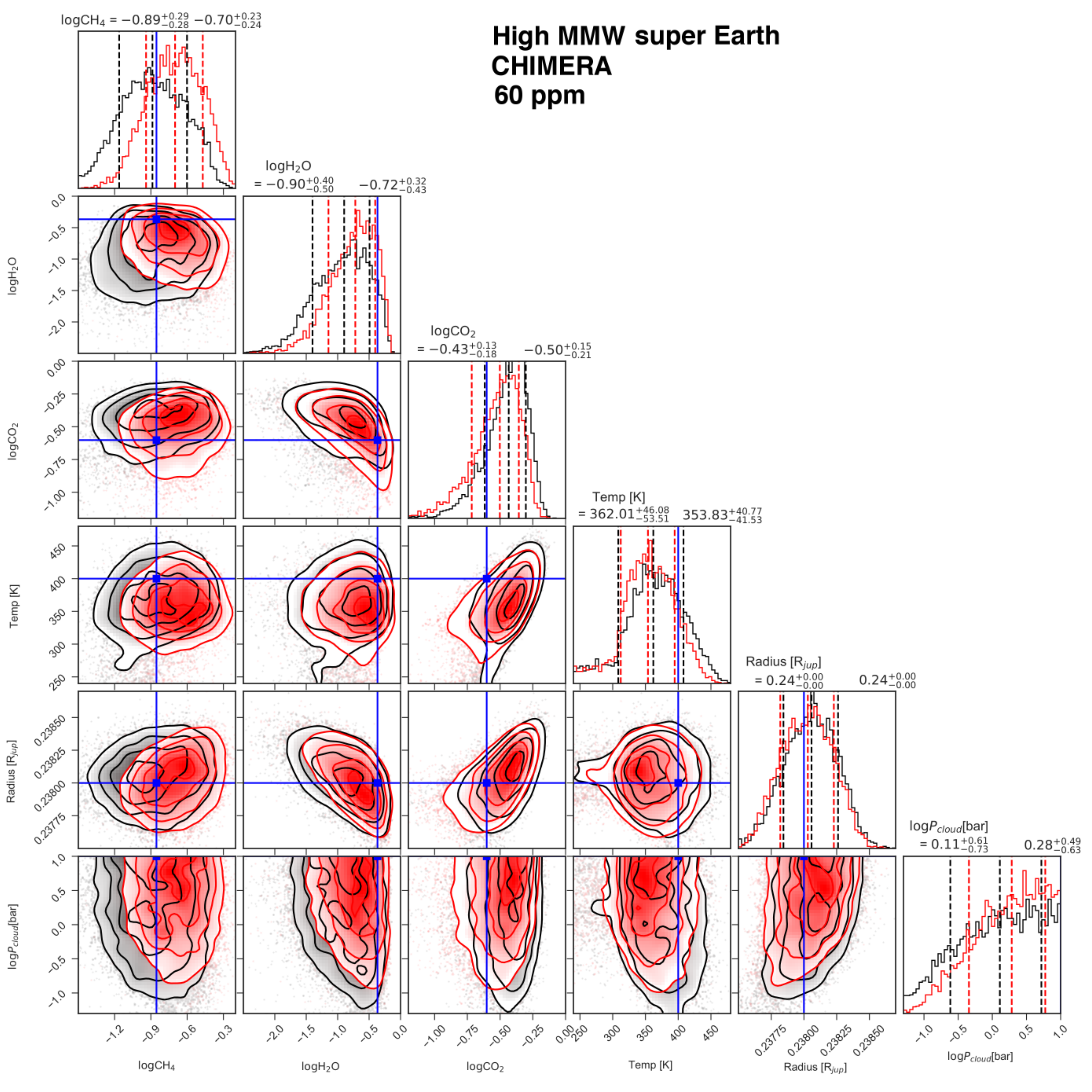}
\caption{As Figure~\ref{planet4_chimera_30} but for 60 ppm. \label{planet4_chimera_60}}
\end{figure*}

\begin{figure*}
\centering
\includegraphics[width=1.0\textwidth]{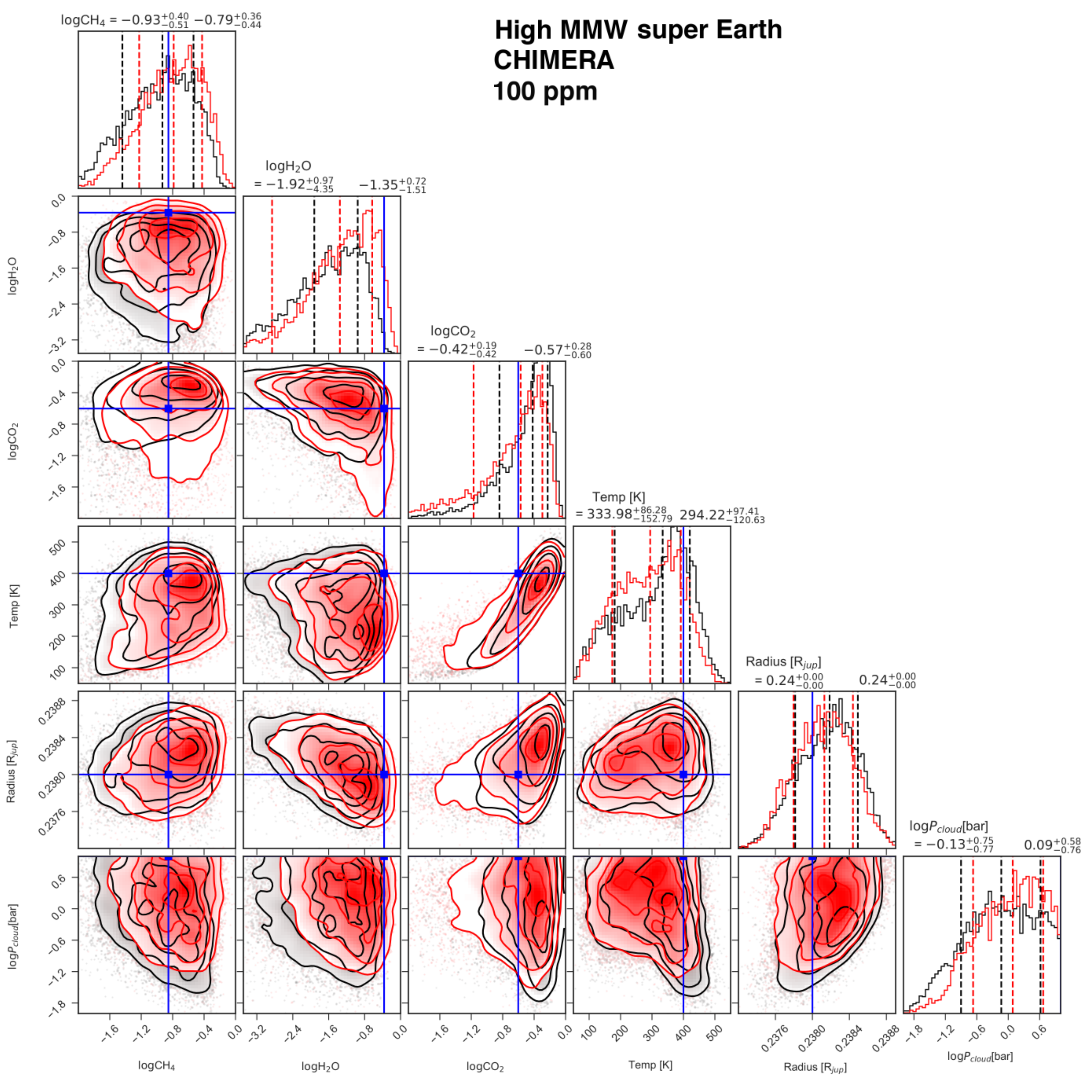}
\caption{As Figure~\ref{planet4_chimera_30} but for 100 ppm. \label{planet4_chimera_100}}
\end{figure*}

A direct comparison of retrieval results for this planet along the lines of those presented for planets 0--3 is not possible, due to differences in the retrieval process where spectrally active gases form a substantial fraction of the atmosphere. In these cases, \nemesis and \chimera both work by forcing the volume mixing ratios of all atmospheric gases to sum to 1. Figures~\ref{planet4_chimera_30}--~\ref{planet4_chimera_100} show retrievals for \chimera on \taurex and \nemesis spectra. This effectively breaks the complex degeneracy between gas abundance, temperature and mean molecular weight, for high signal-to-noise cases, but can only be relied on in cases where the modeller is confident that all significant gas species are included in the model, including non-spectrally active species. As explained above, there are likely to be scenarios where this is not the case, and other approaches may be needed. For spectra with lower signal-to-noise, it may only be possible to achieve upper or lower limits for some gases (Figure~\ref{planet4_chimera_100}). 

\begin{figure*}
\centering
\includegraphics[width=1.0\textwidth]{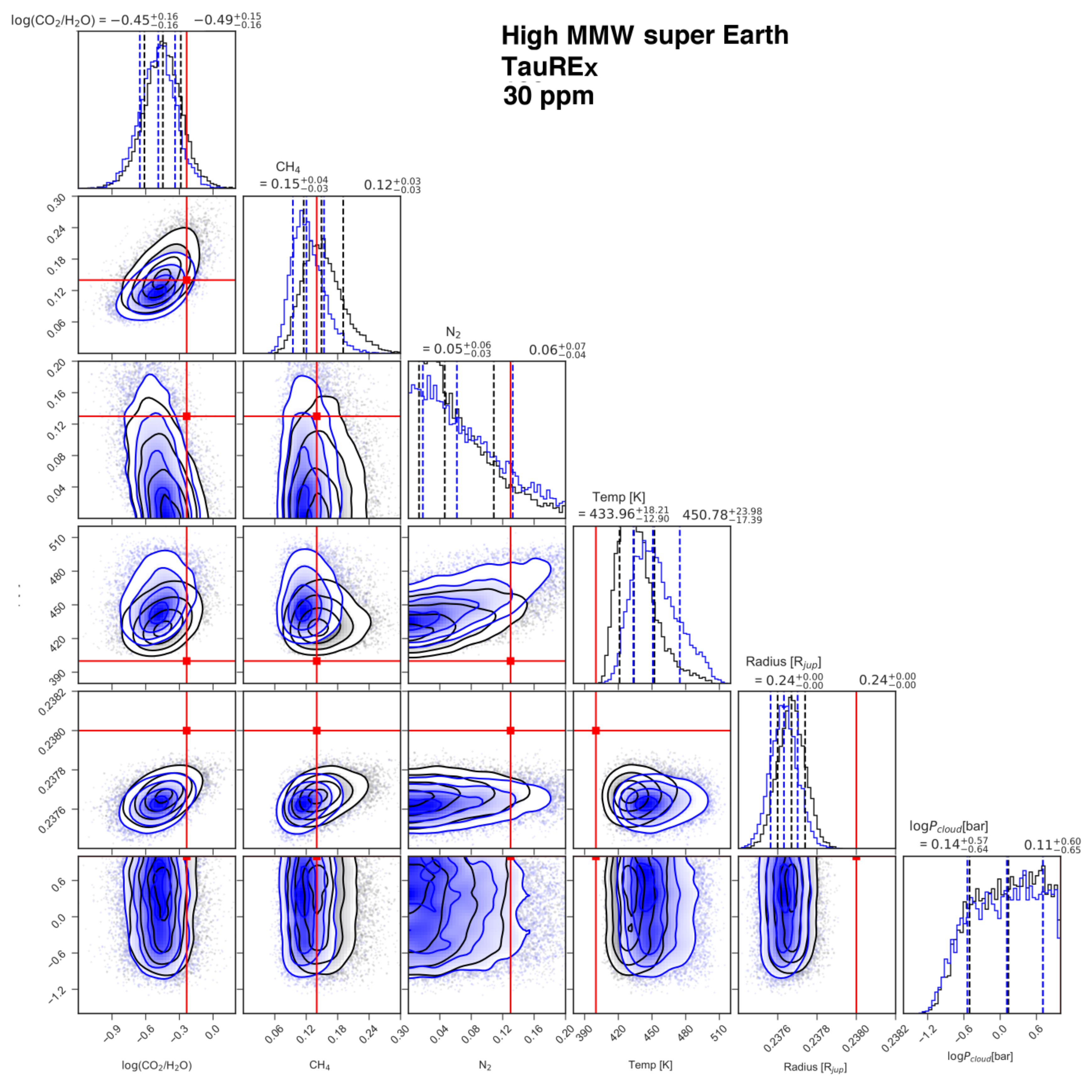}
\caption{As Figure~\ref{planet4_chimera_30} but using \taurex to retrieve \nemesis (black) and \chimera spectra (blue). Note that for \taurex the ratio of CO$_2$ over H$_2$O is retrieved rather than the absolute abundances, and the N$_2$ abundance is retrieved explicitly.  \label{planet4_taurex_30}}
\end{figure*}

\begin{figure*}
\centering
\includegraphics[width=1.0\textwidth]{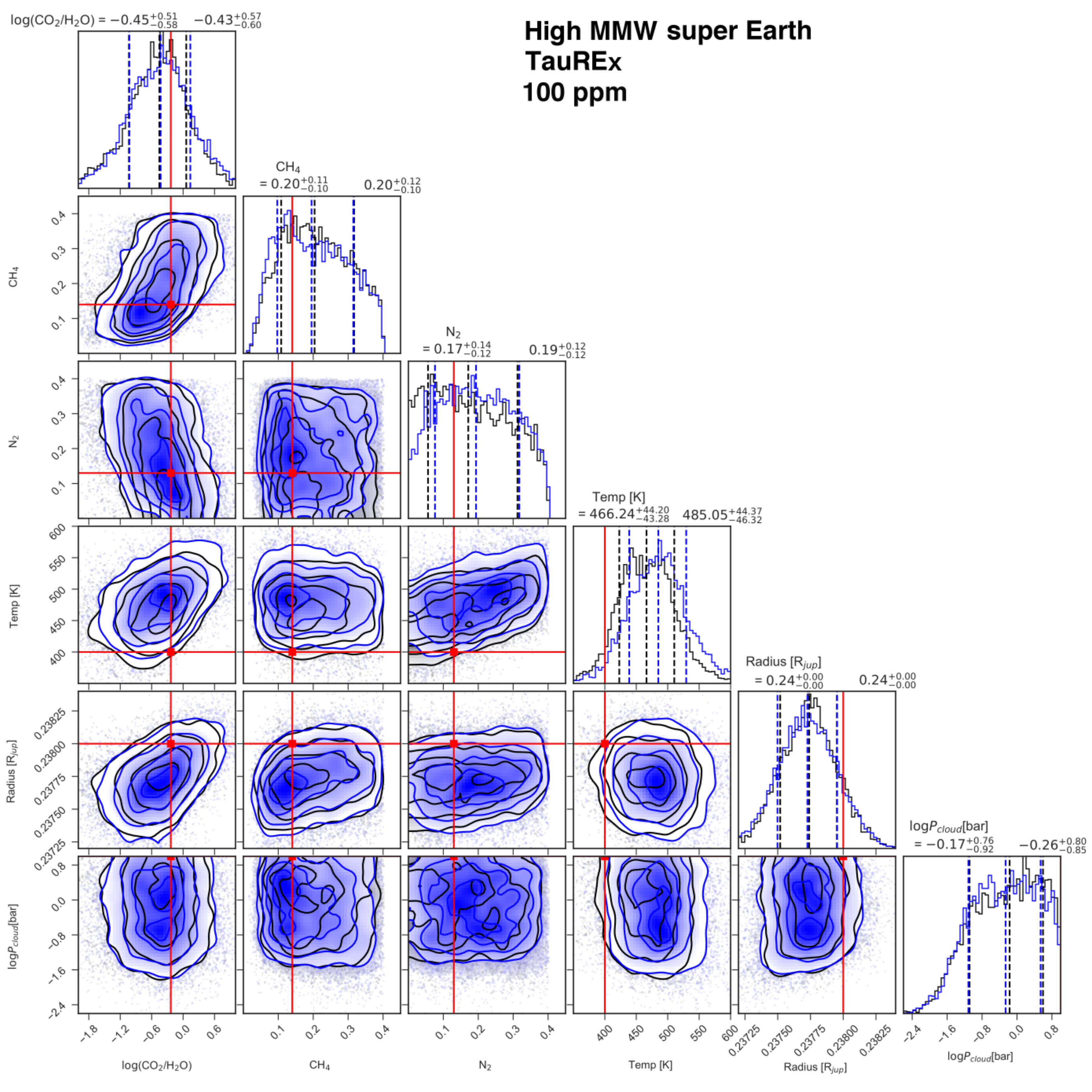}
\caption{As Figure~\ref{planet4_taurex_30} but for 100 ppm. \label{planet4_taurex_100}}
\label{lastpage}
\end{figure*}

Instead of forcing the volume mixing ratios to sum to 1, ~\taurex can be set up to retrieve ratios of gases rather than absolute abundances. Retrieving the ratio of CO$_2$/H$_2$O instead of the absolute abundances, and directly retrieving the N$_2$ abundance ratio via its effect on the mean molecular weight, resolves the problem of forcing the gases to add to unity (Figure~\ref{planet4_taurex_30}). This approach recovers the correct abundance ratio for CO$_2$/H$_2$O as well as the correct absolute abundances for CH$_4$ and N$_2$; however, the retrieved temperature is higher than the input by $>$ 1$\sigma$ for the 100 ppm case (Figure~\ref{planet4_taurex_100}) and $>$ 2$\sigma$ for the other cases, with similar errors in the opposite direction for the radius, illustrating that being agnostic about whether or not the model contains all relevant gases introduces further degeneracy to the problem. 

\section{Discussion}
The synthetic retrievals presented in Section~\ref{retrievals} provide a useful analogy for the likely challenges that will be faced when interpreting spectra obtained by JWST. Whilst they clearly demonstrate that it is possible to recover information about chemistry, temperature and clouds from low-resolution exoplanet transmission spectra, they also illustrate the limits of retrieval capability. 

Current state-of-the-art retrievals utilise Normal likelihood functions. This makes the implicit assumption that observed noise is photon noise dominated and instrument systematics are negligible  For current space-based measurements, such an assumption may not be correct but the currently available signal-to-noise ratios (SNRs) make noise characterisation often infeasible. For high SNR JWST observations, this is unlikely to be the case; systematic sources of error will become important and measurable, and must be formally accounted for in retrieval likelihoods. As well as instrument systematics, other unknowns will also become important and will be necessary to mitigate; the spectral effects of starspots and faculae (e.g. \citealt{rackham18}) could introduce systematic errors into the planetary transmission spectrum. 

Potentially the most significant source of systematic error is model unknowns. We have demonstrated that there are small discrepancies between the forward models used here that are significant enough to result in differences between solutions where the error envelope is small. In this case, models and synthetic observations have the same inputs and level of complexity; when performing retrievals on real data, we should remain aware that retrieval models are simplified representations of complex atmospheric phenomena. Inevitably, an important source of error is the fact that our models are incomplete. For example, we often assume molecular abundances are both vertically and spatially homogeneous; we simplify cloud by parameterising scattering properties using a power law, or even treating it as a grey absorber as we have done here; and we make a choice about which gas species to include in our model atmospheres, meaning we may omit important sources of opacity.   

Whilst it is important to bear these facts in mind, they do not negate the usefulness of retrieval methods using simple parameterised models. Especially in an area such as exoplanet science, where we are currently pushing the boundaries of our knowledge, more detailed physical models also suffer from incompleteness. Retrieval methods, since they are relatively agnostic to prior assumptions, are therefore a necessary tool for discoveries of the unexpected. 

\section{Conclusions}
We have here presented the first systematic comparison of exoplanet retrieval suites. This covers only a limited subset of models and planet conditions, but serves to illustrate the challenges of ensuring consistency between different approaches. We hope the results presented will encourage the exoplanet retrievals community as a whole to continue with such comparative efforts. To this end, we have made our entire suite of forward models and retrieval results available to serve as a benchmark.

We note that the apparently small differences in our forward models sometimes translate into differences in retrieved results outside of the 1-$\sigma$ range, especially in cases where the photon noise level is low (30\,ppm). This can be viewed as an analogy for systematic instrumental and astrophysical noise in real observations, which will become critical with future telescopes such as \textit{JWST} for which the noise floor is $\sim$30\,ppm. It illustrates the importance of awareness of these potential sources of bias, and also the necessity to characterise them to counter the effects.  

Our study also emphasises the importance of accurate molecular and atomic linelists. The linelists used in these calculations were the most complete available at the time the work was performed, but these databases are constantly being improved and updated. Complete linelists are critical for accurate determination of gas abundances.

\section*{Contributions and Acknowledgements}
This paper is the result of several years of collaborative effort between all co-authors; names are therefore listed alphabetically and position in the author list is not a reflection of relative contribution. The original idea was conceived in discussion with JKB, MRL, MR and IPW. JKB set up the initial \nemesis simulations and performed the most recent retrievals. The upgrade to MultiNest and the bulk of \nemesis modelling and retrievals were performed by RG. \taurex simulations and retrievals were performed by QC, MR and IPW. All \chimera work was undertaken by MRL. JKB prepared the manuscript. 

JKB is supported by a Royal Astronomical Society Research Fellowship. This project has also received funding from the European Research Council (ERC) under the European Union's Horizon 2020 research and innovation programme (grant agreement No 758892, ExoAI; No. 776403, ExoplANETS A) and under the European Union's Seventh Framework Programme (FP7/2007-2013)/ ERC grant agreement numbers 617119 (ExoLights). Furthermore, we acknowledge funding by the Science and Technology Funding Council (STFC) grants: ST/K502406/1, ST/P000282/1, ST/P002153/1 and ST/S002634/1. MRL acknowledges support from  NASA Exoplanet Research Program award NNX17AB56G.
JKB and RG thank Patrick Irwin for the use of \nemesis and for his tireless assistance with debugging. 
We thank Dan Foreman-Mackey for the use of the corner.py routine, which is available to download on GitHub: https://github.com/dfm/corner.py. We are grateful to the Center for Open Science for their provision of the free Open Science Framework data hosting service. We also thank the anonymous referee for suggestions that improved the clarity of the manuscript.

\bibliographystyle{mnras.bst}
\bibliography{bibliography.bib}

\end{document}